\documentclass[a4paper,fleqn,usenatbib,useAMS]{mnras}

\usepackage{graphicx}	
\usepackage{amsmath}	
\usepackage{amssymb}	
\usepackage{multicol}        
\usepackage{bm}		
\usepackage{pdflscape}	

\usepackage[T1]{fontenc}
\usepackage{ae,aecompl}

\usepackage{txfonts}

\title[GAMA void galaxies]{Galaxy And Mass Assembly (GAMA): The Bright Void Galaxy Population in the Optical and Mid-IR}

\author[S. J. Penny et al.]{
S.~J.~Penny,$^{1,2,3}$\thanks{E-mail: samantha.penny@port.ac.uk (SJP)}
M. J. I. Brown,$^{1,2}$ 
K. A. Pimbblet,$^{4,1,2}$ 
M. E. Cluver,$^{5}$  
\newauthor D. J. Croton,$^{6}$  
M. S. Owers,$^{7,8}$ 
R. Lange,$^{9}$
M. Alpaslan,$^{10}$ 
I. Baldry,$^{11}$ 
\newauthor J. Bland-Hawthorn,$^{12}$ 
S. Brough,$^{7}$   
S. P. Driver,$^{9,13}$ 
B. W. Holwerda,$^{14}$ 
A. M. Hopkins,$^{7}$ 
\newauthor T. H. Jarrett,$^{15}$   
D. Heath Jones,$^{8}$
L. S. Kelvin$^{16}$
M. A. Lara-L\'opez,$^{17}$ 
 J. Liske,$^{18}$
\newauthor A. R. L\'opez-S\'anchez,$^{7,8}$  
J. Loveday,$^{19}$ 
M. Meyer,$^{9}$
P. Norberg,$^{20}$
A. S. G. Robotham,$^{9}$ 
\newauthor and M. Rodrigues$^{21}$
\\
$^{1}$School of Physics, Monash University, Clayton, Victoria 3800, Australia\\ 
$^{2}$Monash Centre for Astrophysics, Monash University, Clayton, Victoria 3800, Australia\\
$^3$Institute of Cosmology and Gravitation, University of Portsmouth, Dennis Sciama Building, Burnaby Road, Portsmouth, PO1 3FX, UK \\ 
$^4$ Department of Physics and Mathematics, University of Hull, Cottingham Road, Kingston-upon-Hull HU6 7RX, UK\\
$^5$ University of the Western Cape, Robert Sobukwe Road, Bellville, 7535, South Africa\\
$^{6}$ Centre for Astrophysics and Supercomputing, Swinburne University of Technology, Hawthorn, Victoria 3122, Australia\\
$^7$ Australian Astronomical Observatory, PO Box 915, North Ryde, NSW 1670, Australia\\
$^{8}$ Department of Physics and Astronomy, Macquarie University, NSW 2109, Australia\\
$^9$ ICRAR, The University of Western Australia, 35 Stirling Highway, Crawley, WA 6009, Australia\\
$^{10}$ NASA Ames Research Centre, N232, Moffett Field, Mountain View, CA 94035, United States\\
$^{11}$ Astrophysics Research Institute, Liverpool John Moores University, IC2, Liverpool Science Park, 146 Brownlow Hill, Liverpool, L3 5RF\\
$^{12}$ Sydney Institute for Astronomy, School of Physics A28, University of Sydney, NSW 2088, Australia\\
$^{13}$ Scottish Universities' Physics Alliance (SUPA), School of Physics and  Astronomy, University of St Andrews, North Haugh, St Andrews, KY16 9SS, UK\\
$^{14}$ University of Leiden, Sterrenwacht Leiden, Niels Bohrweg 2, NL-2333 CA Leiden, The Netherlands\\
$^{15}$ University of the Cape Town, Astronomy Department, Rondebosch, 7701, South Africa \\
$^{16}$ Institut f{\"u}r Astro- und Teilchenphysik, Universit{\"a}t Innsbruck, Technikerstra{\ss}e 25, 6020 Innsbruck, Austria\\
$^{17}$ Instituto de Astronom\'ia, Universidad Nacional Aut\'onoma de M\'exico, A.P. 70-264, 04510 M\'exico, D.F., M\'exico\\
$^{18}$ Hamburger Sternwarte, Universit{\"a}t Hamburg, Gojenbergsweg 112, 21029 Hamburg, Germany\\
$^{19}$ Astronomy Centre, University of Sussex, Falmer, Brighton BN1 9QH, UK\\
$^{20}$ ICC \& CEA, Department of Physics, Durham University, South Road, Durham DH1 3LE, UK\\
$^{21}$ GEPI, Observatoire de Paris, CNRS, University Paris Diderot ; 5 Place Jules Janssen, 92195 Meudon, France
}

\date{Accepted XXX. Received YYY; in original form ZZZ}

\pubyear{2015}

\begin{document}
\label{firstpage}
\pagerange{\pageref{firstpage}--\pageref{lastpage}}
\maketitle

\begin{abstract}
We examine the properties of galaxies in the Galaxies and Mass Assembly (GAMA) survey located in voids with radii $>10~h^{-1}$~Mpc.  
Utilising the GAMA equatorial survey, 592 void galaxies are identified out to $z\approx0.1$ brighter than $M_{r} = -18.4$, our magnitude completeness limit.  
Using the $W_{\rm{H\alpha}}$ vs. [N\textsc{ii}]/H$\alpha$ (WHAN) line strength diagnostic diagram, we classify their spectra as star forming, AGN, or dominated by old stellar populations.
For objects more massive than $5\times10^{9}$~M$_{\odot}$, we identify a sample of 26 void galaxies with old stellar populations classed as passive and retired galaxies in the WHAN diagnostic diagram, else they lack any emission lines in their spectra. 
When matched to \textit{WISE} mid-IR photometry, these passive and retired galaxies exhibit a range of mid-IR colour, with a number of void galaxies exhibiting  $[4.6]-[12]$ colours inconsistent with completely quenched stellar populations, with a similar spread in colour seen for a randomly drawn non-void comparison sample.
We hypothesise that a number of these galaxies host obscured star formation, else they are star forming outside of their central regions targeted for single fibre spectroscopy. 
When matched to a randomly drawn sample of non-void galaxies, the void and non-void galaxies exhibit similar properties in terms of optical and mid-IR colour, morphology, and star formation activity, suggesting comparable mass assembly and quenching histories. 
A trend in mid-IR $[4.6]-[12]$ colour is seen, such that both void and non-void galaxies with quenched/passive colours $<1.5$ typically have masses higher than $10^{10}$~M$_{\odot}$, where internally driven processes play an increasingly important role in galaxy evolution.  
\end{abstract}




\section{Introduction}

Redshift surveys reveal a remarkable amount of structure, with the majority of galaxies located in groups, clusters, and along the filaments linking these massive structures \citep[e.g.][]{1983ApJS...52...89H,2000AJ....120.1579Y,2001MNRAS.328.1039C}. 
However, it is the voids that represent most of the volume of the Universe. 
Spanning tens of megaparsecs, these void regions are extremely under-dense, with a galaxy density less than $20$~per~cent of the cosmic mean.
These voids are not empty, and contain a sizeable galaxy population \citep[e.g.][]{2004ApJ...617...50R, 2008MNRAS.386.2285C, 2012MNRAS.421..926P, 2012AJ....144...16K}. 
Nevertheless, $\Lambda$ Cold Dark Matter ($\Lambda$-CDM) cosmology predicts more dark matter haloes in voids than we currently observe galaxies \citep{2001ApJ...557..495P,2009MNRAS.395.1915T,2010Natur.465..565P}, and a better understanding the evolution of these galaxies will help us better understand this discrepancy. 
Void galaxies of all masses are therefore ideal objects in which to examine intrinsic vs. extrinsic effects on galaxy evolution, having formed far from the nearest cluster mass dark matter halo. 

Galaxies can be broadly divided into two categories based on their optical colours: blue cloud galaxies with ongoing star formation, and red sequence galaxies with colours consistent with a non-star forming stellar population, though there is colour overlap between these two populations \citep[e.g.][]{2014arXiv1408.5984T}. 
Clearly at some point in their  histories all galaxies were star forming- it is how and where the quenching of star formation takes place that is one of the most active areas of research today. 
These quenching mechanisms fall into two classes:  intrinsic (driven by galaxy mass) vs. extrinsic (environmentally driven) evolution. 

Galaxies follow a strong morphology-density relation, such that early-type galaxies with little or no star formation are primarily found in high density regions of the Universe such as groups and clusters, while late-type galaxies dominate in low density environments such as the field \citep{1980ApJ...236..351D,2005ApJ...620...78S,2007ApJ...658..898P}. 
Similar environmental trends are seen for colour and star formation, such that red galaxies with no star formation favour high density environments \citep{2002MNRAS.334..673L,2004MNRAS.353..713K,2008MNRAS.383..907B,2009MNRAS.393.1324B}. 
This is reflected in the colour-magnitude relation, such that a strong red-sequence is observed for cluster members, with the fraction of red galaxies  decreasing out to the field. 
The red cluster galaxies typically have reduced star formation rates relative to galaxies in the field \citep[e.g.][]{1997ApJ...488L..75B,1998ApJ...504L..75B}, along with little to no gas \citep[e.g.][]{2007A&A...474..851D}.

At a na{\"i}ve first glance, it appears that environment is the clear driver in galaxy evolution: when a galaxy enters a high density region of the Universe (e.g. a cluster), it is stripped of star forming material, and it ceases star formation.
Processes such as galaxy harassment \citep{1996Natur.379..613M}, viscous stripping 
\citep[e.g.][]{2008MNRAS.388.1245R,2013ApJ...765...93C}, strangulation \citep{1980ApJ...237..692L}, tidal stripping, and ram-pressure stripping \citep{1972ApJ...176....1G}, can all disrupt galaxies and remove their fuel for star formation \citep[e.g.][]{2012MNRAS.424..232W}. 
However, it has been observed that the current level of star formation (or colour) in a galaxy is more strongly correlated to galaxy mass than local environmental density \citep[e.g.][Alpaslan et al., sumitted]{2007MNRAS.381....7H,2012MNRAS.423.3679W}, though conflicting results are also found \citep[e.g.][]{2004ApJ...615L.101B}.

While the above environmental processes certainly play an important role in quenching star formation in the satellite galaxy population, the situation becomes unclear for central, high mass galaxies. 
Galaxies with high stellar masses are typically red in all environments. 
Indeed, high mass, isolated, early-type galaxies exist, with examples in the nearby Universe including NGC~3332, NGC~5413 and IC~1156 \citep{2001AJ....121..808C}. 
These galaxies are optically red, with no evidence of star formation (such as H$\alpha$ emission) in their nuclear/central spectra. 
Mass-quenching through processes such as AGN feedback and the virial shock heating of infalling cold gas are important at high galaxy mass as these processes become more efficient \citep[e.g.][]{2013MNRAS.428.3306W}. 
Gas-supply can also be heavily depleted during episodes of star formation, triggered during merger episodes such as minor mergers with satellite galaxies.

Separating these extrinsic and intrinsic quenching mechanisms is not trivial. 
By examining galaxies in voids, we are largely excluding the environmental effects present in groups and clusters. 
Major mergers and interactions between void galaxies are expected to be rare (though not entirely absent, see \citealt{2013AJ....145..120B} for an example of an interacting system in a void).
Like other isolated galaxies, void galaxies are expected to build up their mass  primarily via star formation and minor mergers, resulting in discy morphologies. 
These void galaxies are not entirely cutoff from the cosmic web, linked via tendrils joining them to galaxy filaments \citep{2004MNRAS.350..517S,2008MNRAS.390..408Z,2014MNRAS.440L.106A}, continually supplying them with the fuel for star formation, and increasing the chance of minor mergers.

Examining the void galaxy population is impossible without deep, wide-field redshift surveys. 
Such surveys are vital not only to provide a statistically significant sample of void galaxies, but also to identify the void regions themselves via accurate positional and distance measurements.  
In this paper, we examine void galaxies in the Galaxy and Mass Assembly (GAMA) survey \citep{2009A&G....50e..12D}, utilising optical and infrared data. In addition to data at optical wavelengths provided by the GAMA and SDSS surveys, we furthermore use mid-IR data to search the GAMA void galaxy population for obscured star formation. 
Sensitive to the emission from dust-reprocessed star formation, data from the mid-IR is crucial in establishing if a population of truly passive void galaxies exists  (intrinsic vs. extrinsic evolution).

Uncovering the nature of mass quenching first requires obtaining a population of well-isolated galaxies with no ongoing star formation, spanning the stellar mass range at which the fraction of quenched galaxies increases \citep[$\approx 3\times10^{10}$~M$_{\odot}$;][]{2003MNRAS.341...54K}. 
The most basic way to do this is via the colour-mass relation, selecting galaxies that are passive in terms of their optical colours (they lie on the red sequence), and that do not exhibit spectral features consistent with ongoing star formation, via emission line diagnostics such as the Baldwin, Phillips \& Terlevich \citep[BPT,][]{1981PASP...93....5B,2006MNRAS.372..961K} or $W_{\rm H\alpha}$ vs. [N\textsc{ii}]/H$\alpha$  \citep[WHAN,][]{2011MNRAS.413.1687C} diagrams. 

Examining the stellar populations of these optically passive galaxies in other wavelengths can reveal a different picture to the optical. For example, galaxies residing on the optical red sequence may exhibit a UV excess consistent with star formation in the past 1~Gyr \citep{2005ApJ...619L.111Y,2007ApJS..173..512S,2014MNRAS.437.2521C}. 
The mid-IR provides further evidence that not all early-type galaxies or optically red galaxies are passively evolving. 
\citet{2009MNRAS.392..982C} show that 32~per~cent of early-type galaxies in the Coma Cluster are not passive, and their mid-IR colours lie off the tight  $K_{s} -[16]$ colour sequence traced by quenched galaxies.   
\citet{2013ApJ...767...90K} match quiescent red-sequence galaxies with no H$\alpha$ emission to \textit{Wide-field Infrared Survey Explorer (WISE)} 12~$\mu$m photometry, and find that 55~per~cent of their sample exhibits excess mid-IR emission consistent with star formation in the past 2~Gyr. 
Using a combination of mid-IR and optical colours, we therefore gain a better understanding of a galaxy's stellar population than at optical wavelengths alone, and identify truly passive galaxies in both the void and general galaxy population.   

The aim of this paper is to examine if the mechanisms responsible for the regulation and truncation of star formation differ as a function of large-scale environment. 
We do this by comparing galaxies that reside on the cosmic web in clusters, groups, and filaments, with those that reside in the under-dense voids that separate these structures (though see \citealt{2004MNRAS.350..517S} and \citealt{2014MNRAS.440L.106A} for evidence that void galaxies are, in fact, linked to the cosmic web via tendrils). 
The voids we examine have a typical galaxy density 20~per~cent of the cosmic mean. 
We use the void galaxy population to search for signs of mass quenching in some of the most remote galaxies in the Universe using data from GAMA, including GAMA-\textit{WISE} data from \citet{2014ApJ...782...90C}. 
Using a combination of optical $(u-r)$ and mid-IR colours, along with emission line diagnostics, we find a population of void galaxies with masses $>5\times10^{9}$ and both optical and mid-IR colours consistent with passively evolving stellar populations. 
By selecting these void galaxies, we are largely removing the environmental effects that influence galaxy evolution, and we instead focus on intrinsic evolutionary processes. 
If well-isolated void galaxies with high masses are a passively evolving population, they must follow the same evolutionary pathways as comparable central galaxies in the rest of the Universe (i.e. mass quenching). 

This paper is ordered as follows. The GAMA survey is described in Section~\ref{sec:gama}, with our photometry and stellar mass sources presented in Section~\ref{sec:opphot}. 
Our void galaxy sample, non-void comparison sample, and  completeness limits are presented in Section~\ref{sec:vgals}. 
The colour mass relation for void galaxies is presented in Section~\ref{sec:red}, and the selection of passive galaxies via the WHAN line strength diagnostic diagram is presented in Section~\ref{sec:whan}. 
\textit{WISE} photometry is utilised in Section~\ref{sec:gamawise} to search for ongoing star formation in optically passive galaxies, and to search for evidence of mass quenching in the void population in Section~\ref{sec:irmassquench}.  
The properties of the highest mass void galaxies ($M_{\star} > 10^{10}$~M$_{\odot}$) are examined in Section~\ref{sec:higmass}, and we examine the merger histories of isolated galaxies via a comparison to the Millennium Simulation in Section~\ref{sec:modelmergers}. 
We discuss our results in Section~\ref{sec:discuss}, and conclude in Section~\ref{sec:conclude}.

Throughout this paper we  use $\Omega_{\Lambda} = 0.7$, $\Omega_{M} = 0.3$, and $H_{0} = 70$~km~s$^{-1}$, but for the comparison with \citet{2012MNRAS.421..926P} we use $h$ in our notation to simplify comparison with the prior literature.  
Optical magnitudes are given in the AB system \citep{1983ApJ...266..713O}. 
The \textit{WISE} survey is calibrated to the Vega magnitude system, and mid-IR photometry is therefore presented in the Vega system,  allowing for easy comparison to the literature.

\section{The Galaxy and Mass Assembly Survey}
\label{sec:gama}

The Galaxy and Mass Assembly \textsc{ii} (GAMA \textsc{ii}) survey is a multi-wavelength photometric and spectroscopic survey that covers three equatorial regions centred on $\alpha = 9^{\rmn{h}}$, $\delta = +0.5^\circ$ (G09),  $\alpha = 12^{\rmn{h}}$, $\delta = -0.5^\circ$ (G12), $\alpha = 14.5^{\rmn{h}}$, $\delta = +0.5^\circ$ (G15), along with two non-equatorial regions.  
Each equatorial region covers $12^{\circ} \times 5^{\circ}$. 
The spectroscopic survey targets galaxies to $r <19.8$ in the G09, G12 and G15 equatorial regions, to a high redshift completeness of $>98$~per~cent (Liske et al., submitted). 
The majority of spectra in GAMA were taken using the AAOmega spectrograph \citep{Saunders04} on the 3.9~m Anglo-Australian Telescope at the Siding Spring Observatory, with this main spectroscopic sample supplemented by data from other surveys such as the Sloan Digital Sky Survey (SDSS) data release 7 \citep{2009ApJS..182..543A}. 
The reduction and analysis of the AAOmega spectra is described in \citet{2013MNRAS.430.2047H}. 

Further details of the GAMA survey characteristics are given in \citet{2011MNRAS.413..971D} and \citet{2015MNRAS.452.2087L}, the survey input catalogue is described in \citet{2010MNRAS.404...86B}, and the tiling algorithm for positioning the AAOmega spectrograph fibres is described in \citet{2010PASA...27...76R}. 
The reduction and analysis pipeline for the AAOmega spectrograph is described in \citet{2013MNRAS.430.2047H}, with the automated redshift pipeline described in \citet{2014MNRAS.441.2440B}.  
The optical photometry utilised in this work was obtained from SDSS imaging in the \textit{u, g, r, i, z} bands as described in \citet{2011MNRAS.412..765H}. 
Additional data products used in this work are the stellar mass catalogue \citep{2011MNRAS.418.1587T}, the GAMA environment measures catalogue \citep{2013MNRAS.435.2903B},  the GAMA line strength catalogue \citep{2013MNRAS.433.2764G},  and the GAMA-\textit{WISE} catalogue \citep{2014ApJ...782...90C}.
 We also utilise the GAMA Galaxy Group Catalogue \citep{2011MNRAS.416.2640R} to remove interloper high peculiar velocity objects from the void regions, and examine if massive void galaxies are the central galaxies in their haloes, else if they reside in pairs. 
 We only use sources with reliable redshifts (GAMA redshift quality flag $nQ \geq 3$, Liske et al. 2015) in this work to ensure an accurate determination of environment for all galaxies.

\subsection{Photometry and stellar masses}
\label{sec:opphot}

Stellar masses for the GAMA \textsc{ii} survey are provided by the catalogue of \citet{2011MNRAS.418.1587T} with an update to this catalogue to include GAMA \textsc{ii} galaxies.  The catalogue furthermore provides  absolute magnitudes and colours for GAMA galaxies out to $z = 0.65$, $k$-corrected to $z=0$. 
The values provided in the stellar mass catalogue are derived via stellar population synthesis modelling of the galaxy's broadband photometry, via comparison to Bruzual \& Charlot (2003) stellar population synthesis models, assuming a Chabrier initial mass function. A Calzetti et al. (2000) dust curve is assumed. For full details of these stellar mass calculations, see \citet{2011MNRAS.418.1587T}.

The values in this catalogue are calculated using aperture photometry (\textsc{sextractor} \texttt{auto} photometry), which may miss a significant fraction of a galaxy's light. 
We therefore apply an aperture correction from the GAMA stellar mass catalogue to integrated values such as stellar mass and absolute magnitude, to account for flux/mass that falls beyond the aperture radius used for SED matching. 
The GAMA stellar mass catalogue provides the linear ratio between each object's $r$-band aperture flux and the total $r$-band flux from a S\'ersic profile fitted out to 10~$R_{e}$ \citep{2011MNRAS.418.1587T,2012MNRAS.421.1007K}. 

This flux ratio allows us to clean our samples of galaxies with spurious masses which are frequently blended with neighbouring galaxies, else their photometry and colours are strongly affected by nearby bright stars.  
We select galaxies where the magnitude correction is $ -0.2 < mag_{corr} < 0.1$ to ensure we only include galaxies with reliably measured stellar masses and colours in our void and non-void comparison samples. 
This unequal cut is applied as a galaxy with less flux in its S\'ersic profile out to 10~$R_{e}$ than in the $r$-band aperture has unreliable photometry. 
The cuts remove 11~per~cent of the void galaxies prior to magnitude and mass cuts. These cuts are essential in defining a sample of galaxies with optically passive colours consistent with negligible star formation. 

\section{Void galaxy identification}
\label{sec:vgals}

\begin{figure*}
\includegraphics[width=0.97\textwidth]{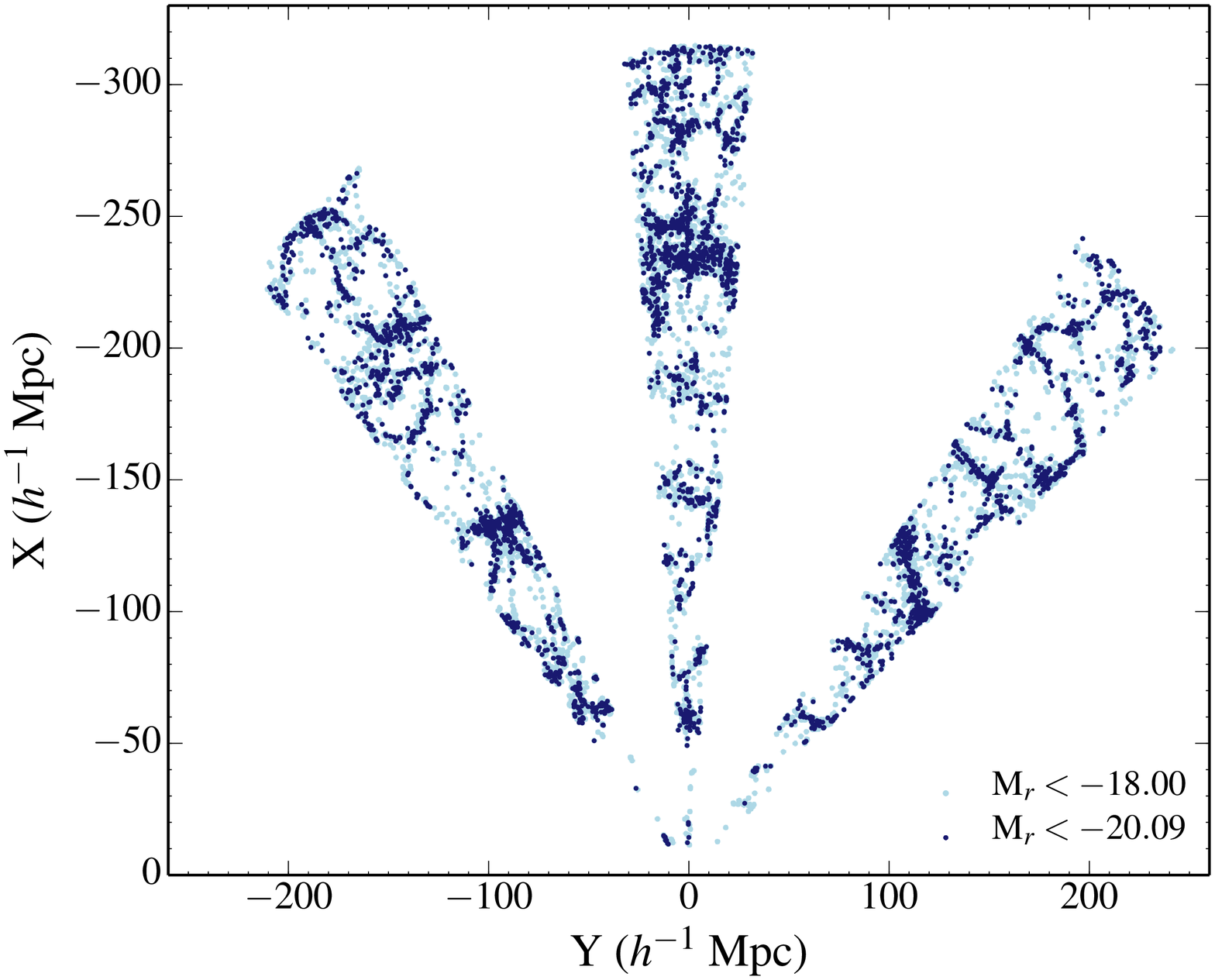}
\caption{The effect of limiting magnitude on void size. Galaxies with $M_{r} < -20.09$ are more strongly clustered than those with $-18.00 < M_{r} < -20.09$, and voids therefore appear larger when using a brighter magnitude cut to define their limits.}\label{voidbounds}
\end{figure*}

\begin{figure*}
\includegraphics[width=0.97\textwidth]{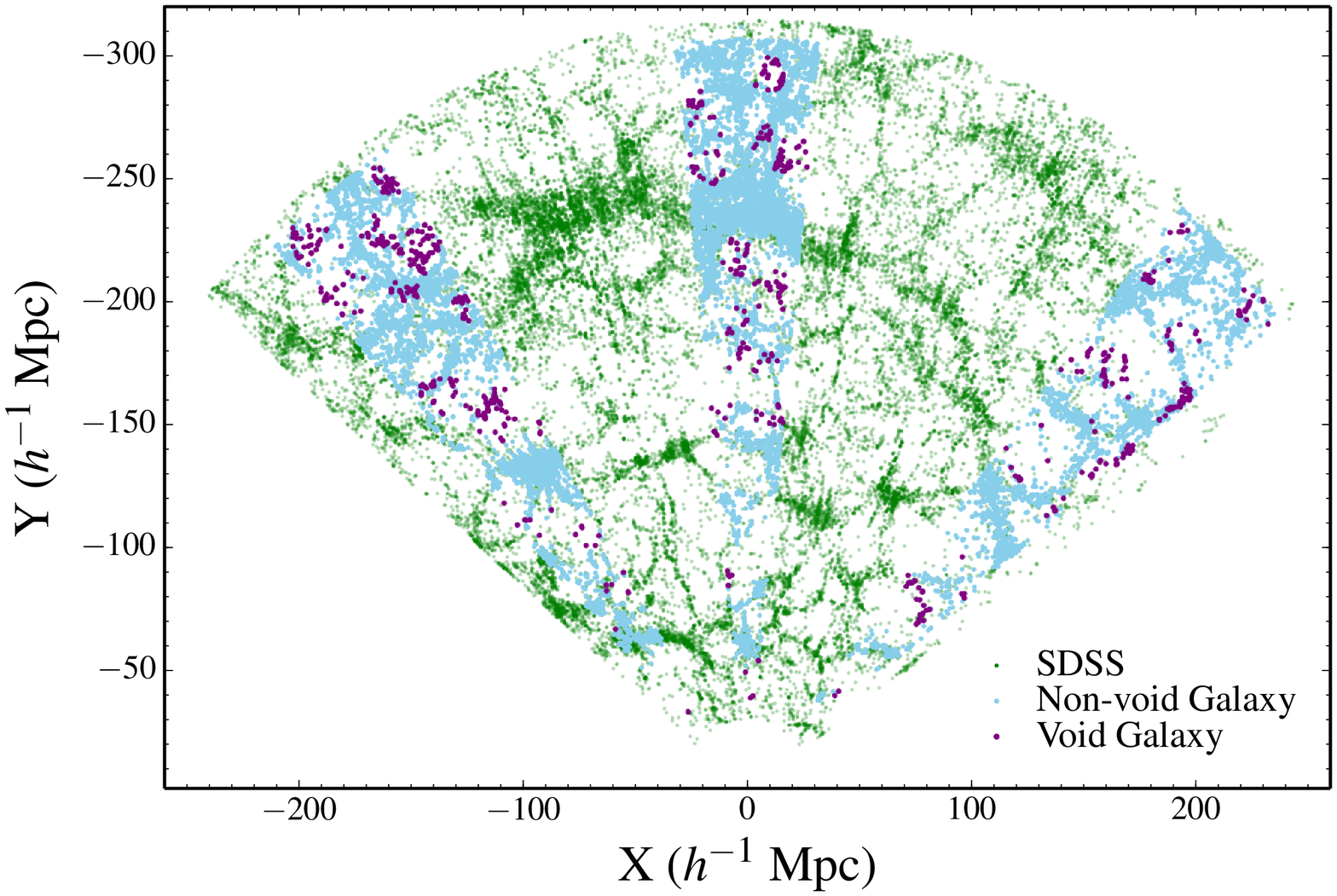}
\caption{Large-scale structure in the GAMA fields (blue points) overlaid on an $10$~$h^{-1}$~Mpc slice of SDSS. The void galaxies identified in this work are shown as purple points. It can be clearly seen that the void galaxies occupy the under-densities in the large scale structure traced by GAMA and SDSS.}\label{las}
\end{figure*}

Cosmological voids are the largest (empty) structures in the Universe, with radii $>10$~$h^{-1}$ Mpc \citep{2012MNRAS.421..926P}.  
The identification of these voids requires large volume surveys with spectroscopic distances determined for its target galaxies. 
Narrow field spectroscopic surveys do not have the volume to completely enclose the large cosmological voids we are interested in for this work.
Due to the narrow-field nature of the equatorial GAMA survey, with each of the three footprints spanning $12^{\circ} \times 5^{\circ}$, a circular void with a radius $R > 10~h^{-1}$~Mpc will not be completely contained within a single GAMA footprint until $z\approx0.06$.
Voids in GAMA may therefore have centres that reside outside of the GAMA survey footprints, and we would miss them and their galaxies if we select voids using GAMA only.  
As targets in the GAMA equatorial regions are selected from the Sloan Digital Sky Survey (SDSS)  \citep{2000AJ....120.1579Y}, we are able to use the large continuous survey area of SDSS ($>7500$~deg$^{2}$) to trace large scale structure in the nearby Universe, and therefore voids which lie in the GAMA survey.  
As a result, our search for void galaxies is confined to the GAMA equatorial (G09, G12, and G15) regions.

The void catalogue of Pan et al. (2012) provides the centres and sizes of voids located throughout the SDSS survey region, including the GAMA equatorial regions out to $z=0.1$. 
The catalogue used heliocentric redshifts. 
The Pan et al. (2012) catalogue uses the Void Finder algorithm (Hoyle \& Vogeley 2002) to separate SDSS into wall and field galaxies, with their field galaxies defined to be all galaxies with a third nearest neighbour distance $d > 6.3$~$h^{-1}$~Mpc. 
This third nearest neighbour distance selects galaxies located in environments with $<20$~per~cent of the mean cosmic density, with densities $<10$~per~cent expected in the void centres.  
The wall galaxies are used to trace large scale structure (filaments, groups, and clusters), and field galaxies are essentially well isolated. 
Void structure is then traced by identifying empty spheres between the wall galaxies. 
The sizes of the void regions are defined by the maximal sphere with radius $R_{void}$ that fills the empty space between the wall galaxies.
We use the void centres defined in this catalogue to find all voids located entirely or partially within the GAMA equatorial survey footprints, and then search for all void galaxies out to three-quarters of the void radius. 
Objects with $z < 0.002$ are excluded to avoid Galactic sources contaminating our sample. 

Only galaxies with magnitudes brighter than $M_{r} = -20.09$ were used by \citet{2012MNRAS.421..926P} in their separation of wall and field galaxies. 
The sizes of voids always depend on the selection criteria used to identify them, and by pushing down the mass/luminosity function, voids can be infilled by low mass galaxies. 
Void size will decrease with a fainter magnitude limit due to the weaker spatial clustering of low mass relative to high mass galaxies \citep{2002MNRAS.332..827N,2011ApJ...736...59Z}. 
We illustrate this in Fig.~\ref{voidbounds}, using a $10~h^{-1}$~Mpc slice though GAMA to highlight large-scale structure. 
The galaxies are split into two luminosity bins: one for galaxies with $M_{r} < -20.09$~mag, and the other for galaxies with $-20.09~{\textrm{mag}} < M_{r} < -18.4$~mag, and their positions in comoving cartesian coordinates plotted.
When we extend this luminosity range down to $M_{r} = -18.4$ (the limiting magnitude at the redshift limit $z=0.1$ of this work), the size of the voids can be seen to decrease as the fainter galaxies are less clustered. 
We therefore only examine void galaxies out to 0.75~$R_{void}$ to remove the less-clustered low mass non-void galaxies from the void galaxy sample. 
By excluding these void-edge galaxies, we ensure that we remove the effects of large-scale environment from our void galaxy sample. 

To the best of our knowledge, the void regions identified in \citet{2012MNRAS.421..926P} are bona fide voids with radii $>10$~$h^{-1}$~Mpc when defined using galaxies brighter than $M_{r} = -20.09$~mag. 
However, \citet{2014MNRAS.440L.106A} identifies fine, low density tendrils that extend into voids defined by bright galaxies. 
We assume that all galaxies examined here belong to their parent void, though they may be connected to regions of higher density by these tendril structures.

We therefore utilise the GAMA environment measures catalogue \citep{2013MNRAS.435.2903B} to examine the local environmental density of the void galaxies. 
To do this, we compare the surface density $\Sigma_{5}$ of the void vs. comparison samples. 
The surface density measure is based on the distance to the 5th nearest neighbour within a velocity cylinder of $\pm1000$~km~s$^{-1}$. 
For the void population, we find a mean surface density $\Sigma_{5} = 0.09 \pm0.055$~Mpc$^{-2}$, vs $\Sigma_{5} = 0.60 \pm0.131$~Mpc$^{-2}$ for the non-void comparison sample. 
We  confirm that the void galaxies examined here are therefore located in extremely under-dense regions of the Universe. 
 
Due to SDSS DR7 spectroscopic incompleteness in the G09 region, this region would be under-dense in the catalogue used to create the list of voids in SDSS.  
We therefore exclude voids and galaxies in the G09 region with Decl. $< 0.0^{\circ}$. 
This incompleteness could result in the identification of artificial voids in regions that in fact contain galaxy groups and clusters. 
We also check our void galaxy sample for the effects of the survey edges. 
There is no preference for red sequence void galaxies to be located at the edges of the survey, i.e. they are evenly distributed.

\subsection{Interloper removal}

Galaxies within groups and clusters can have large peculiar velocities, making them appear at a higher/lower redshift than the over-density in which they reside i.e. the ``fingers of god'' commonly observed for cluster galaxies. 
These peculiar velocities can be sufficient to make a cluster galaxy appear within a void during a simple radial search around a void centre. 
Large groups/clusters will have the highest spread in the peculiar velocities of their member galaxies (e.g. \citealt{2013arXiv1311.4953R}). 

To remove such galaxies from the void sample, we use an update of the GAMA galaxy group catalogue of \citet{2011MNRAS.416.2640R} to identify group galaxies.  
However, we cannot just remove all void galaxies found to reside in groups from our catalogues.  
Simulations predict dark matter substructure and filaments within void regions, consistent with a hierarchical model of galaxy assembly \citep{2009MNRAS.395.1915T,2011ApJ...735..132K,2013MNRAS.435..222R,2013MNRAS.428.3409A}. 
Evidence for this substructure has been found, with a small galaxy group identified in a void consisting of three galaxies embedded in a common H\textsc{i} envelope  \citep{2013AJ....145..120B}, hypothesised to be the assembly of a filament in a void.
To ensure we do not remove such groups, we set a group size limit to separate small void groups from interlopers. 
All void galaxies in groups with $>10$ members are excluded from our final catalogue. 
This limit is selected to keep groups of 2-3 bright galaxies and any satellites. Using this method, 16 galaxies with masses $>10^{9}$~M$_{\odot}$ are identified that belong to a single cluster in G09 with a mass $1.6\times10^{14}$~M$_{\odot}$, and a velocity dispersion $558\pm50$~km~s$^{-1}$ \citep{2011MNRAS.416.2640R}. 
These 16 galaxies are removed from the void galaxy sample. 

The remaining galaxies all reside in groups of 6 or fewer members, else in pairs or isolation. 
For the remaining groups with more than three members, the majority of group members have $M_{r} > -20.09$ and are less massive than $5\times10^{9}$~M$_{\odot}$. 
Due to their low absolute magnitude, these faint galaxies would be absent in the wall/field sample of Pan et al. (2012). 
Indeed, the sum of the luminosities of all members in the only void group with six members (GAMA GroupID 300360) is $M_{r} = -21.4$, fainter than a typical BCG.  
Nevertheless, we retain these galaxies in our sample, as they are well isolated from the filaments that trace large scale structure.

\subsection{Non-void comparison galaxies}
\label{sec:fieldsamp}

To allow the comparison of our void galaxy sample to the general galaxy population, we construct a comparison sample of non-void galaxies in the same volume as the GAMA void galaxies. 
Only galaxies in the G09, G12 and G15 out to $z=0.1024$ (the redshift of the most distant void galaxy) were included in this comparison sample. 
Our non-void comparison sample contains galaxies from a range of environmental density, ranging from clusters through to isolated galaxies in filaments, though void galaxies are removed from the comparison sample. 
These non-void galaxies reside in a range of local galaxy density, ranging from a surface density $\Sigma_{5} < 0.05$~Mpc$^{-2}$ (extremely under-dense), through to over-dense regions with $\Sigma_{5} > 100$~Mpc$^{-2}$ (i.e. clusters). 

As voids are not purely spherical, we furthermore need to avoid the contamination of the comparison sample with void galaxies. 
A void may extend beyond the edges of the maximal sphere used to define it if the void does not have a spherical edge. 
We therefore exclude galaxies to a distance of 1~$h^{-1}$~Mpc from edges of the maximal spheres used to define the void radii (i.e. R$_{void} + 1$~$h^{-1}$~Mpc).
This provides a comparison sample of 14,233 non-void galaxies in the same volume as the void galaxies prior to completeness cuts.
The distribution of the non-void comparison sample is shown Figure~\ref{las} as pale blue dots, with the void galaxies shown as purple dots. 
Also included in Fig.~\ref{las} are the positions of galaxies in SDSS to highlight the presence of large scale structure and voids in the GAMA equatorial regions. 

\subsection{Completeness}
\label{sec:complete}

While we are interested in how the star forming properties of void galaxies vary as a function of galaxy mass, spectroscopic completeness cuts based on galaxy mass alone have a strong luminosity bias as blue, star forming galaxies are brighter at optical wavelengths than passive galaxies for a given stellar mass. 
As such, the low mass end of the galaxy sample will be biased towards star forming galaxies. 
To avoid this luminosity bias in our sample, our completeness cuts placed on our sample are made using \emph{r}-band absolute magnitudes, with aperture corrections applied (see Section ~\ref{sec:opphot}), to ensure no colour bias in our sample.

The spectroscopic limit of GAMA~\textsc{ii} is $r=19.8$, and the redshift completeness at this magnitude limit for the GAMA equatorial regions (G09, G12, and G15) is 98.5~per~cent integrated over all magnitudes, and we therefore use this apparent magnitude as a limit.  

\begin{figure}
\includegraphics[width=0.47\textwidth]{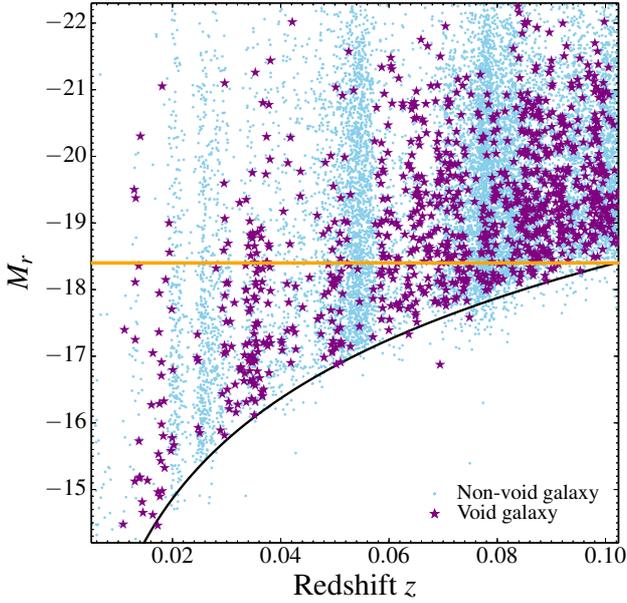}
\caption{The completeness cut used in this work. The horizontal line is the magnitude completeness limit at $M_{r} = -18.4$, and the right-hand y-axis is the redshift limit of this work, which we take to be the redshift of the most distant void galaxy ($z=0.1024$). The black curve is the magnitude selection limit for GAMA ($r=19.8$), which has $\sim98.4$~per~cent spectroscopic completeness integrated over all magnitudes. A number of filler objects targeted with unused 2dF fibres fall below this selection limit. Void galaxies can be seen to occupy the under-densities in the non-void galaxy population. }\label{completeness}
\end{figure}

Our magnitude completeness cuts are shown in Fig.~\ref{completeness}. 
Both the void and non-void galaxy populations are shown, along with the $r=19.8$ spectroscopic limit curve for GAMA. 
To $z=0.1024$, we are complete for galaxies brighter than  $M_{r} =-18.4$ (the horizontal orange line in Fig.~\ref{completeness}). 
We will discuss mass and colour completeness in Sec.~\ref{sec:colmass}.

Low surface brightness galaxies are known to prefer a lower density environment to high surface brightness galaxies \citep[e.g.][]{2009A&A...504..807R,2011ApJ...728...74G}. We do not apply completeness cuts to ensure surface brightness completeness, so we are likely incomplete for low surface brightness objects.  

This magnitude limit of  $M_{r} =-18.4$ excludes dwarf galaxies, which have stellar masses $<10^{9}$~M$_{\odot}$, and are typically fainter than $M_{r} = -18$.  
We are unable to use dwarf  galaxies in our study of mass quenching in void regions. 
However, \citet{2012ApJ...757...85G} show that field dwarf galaxies with no active star formation are extremely rare, comprising $<0.06$~per~cent of galaxies fainter than $M_{r} = -18$ in SDSS Data Release 8. 
No quenched isolated dwarf galaxies with stellar masses $<1.0\times10^{9}$~M$_{\odot}$ are found in their study, so including these faint galaxies is not vital for a study of mass quenching in low density environments.

We furthermore place an upper limit mass limit of $5\times10^{11}$~M$_{\odot}$ on our void and comparison samples. 
Galaxies in our samples with stellar masses higher than this upper limit are typically found to be blended with other sources, or have large background gradients, resulting in their calculated masses being unreliable.  
While galaxies more massive than this exist, we do not expect to find such galaxies in voids, as major mergers between the massive galaxies required to form them are unlikely in these extremely under-dense regions of the Universe.

In  Fig.~\ref{completeness}, it can be clearly seen in that  void galaxies occupy under-densities in the galaxy redshift distribution. 
This is less clear at higher $z$, as the thickness of the wedge increases, sampling a wider range of galaxy density, where voids and groups/clusters may overlap in this plot. 
The number of void galaxies in each GAMA equatorial region for the mass limited sample are shown in Table~\ref{distab}.

\begin{table}
\caption{The distribution of void galaxies with $M_{r} =  -18.4$ over the three equatorial GAMA fields. Due to SDSS spectroscopic incompleteness in the G09 region, we remove void galaxies in this region with $\delta < 0^{\circ}$ to remove the effect of ``false'' voids appearing in this under-sampled region.}\label{distab}
\begin{center}
\begin{tabular}{p{0.8cm} p{1.5cm} p{1.5cm} p{0.7cm} p{1.1cm}}
\hline
Region & $\alpha$ range & $\delta$ range & Void & Non-void \\
 & (J2000.00) & (J2000.00) & & \\
\hline
G09 & $129.0^{\circ}$:$141.0^{\circ}$ & $0.0^{\circ}$:$+3.0^{\circ}$ & 144 &1331 \\
G12 & $174.0^{\circ}$:$186.0^{\circ}$ & $-3.0^{\circ}$:$+2.0^{\circ}$ & 185 & 3861\\
G15 & $211.5^{\circ}$:$223.5^{\circ}$ & $-2.0^{\circ}$:$+3.0^{\circ}$  & 263 & 2626\\
\hline
\end{tabular}
\end{center}
\label{default}
\end{table}%

\section{The colour-mass relation}
\label{sec:colmass}

The colour-magnitude/mass diagram is one of the most basic diagnostic tools when studying galaxy evolution, used to search for trends in galaxy colour with respect to galaxy mass or luminosity. 
Galaxies follow an approximately bimodal colour distribution, splitting into the blue cloud and red sequence, dependent on whether they are currently forming stars.  
Here, we use the $(u-r)$ colour-mass relation to compare the void sample and non-void comparison sample defined in Sec.~\ref{sec:vgals}.

\begin{figure*}
\includegraphics[width=0.98\textwidth]{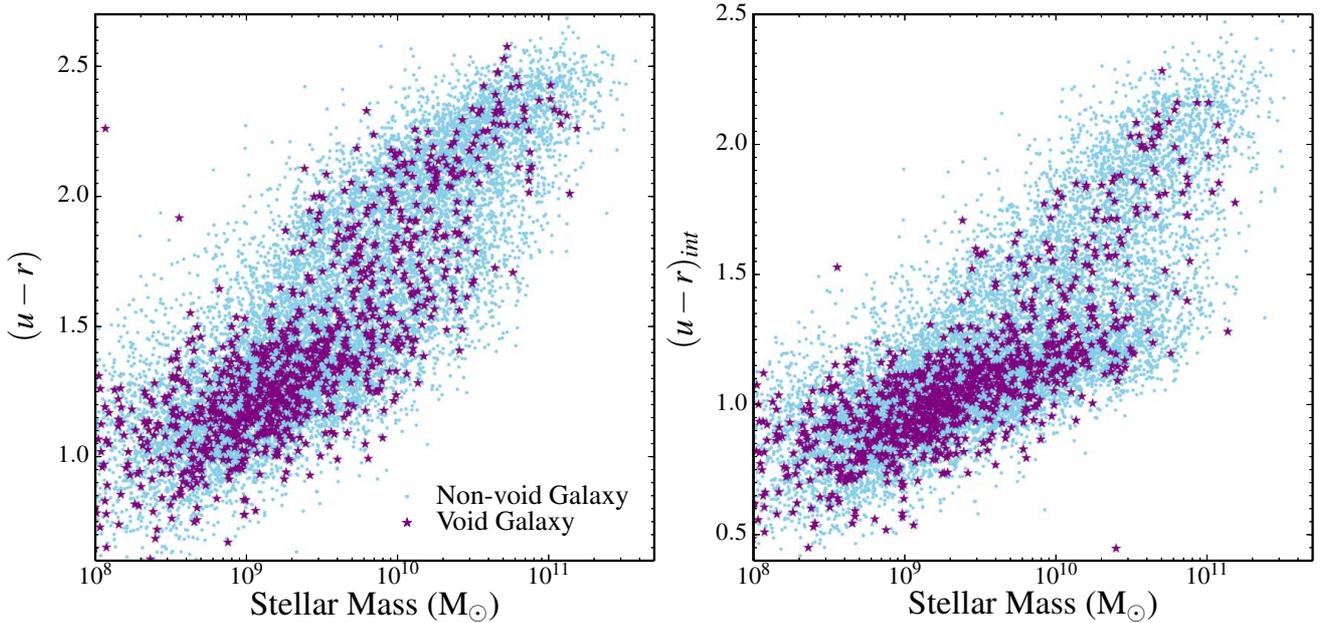}
\caption{Colour-mass relation for GAMA void galaxies (purple stars). The left-hand panel shows rest-frame $(u-r)$ observed colours,  with the rest-frame intrinsic ($u-r$)$_{int}$ colours (corrected for internal dust reddening) shown in the right-hand panel. The colours have been $k$-corrected to $z=0$ in both plots, and corrected for foreground extinction. Also shown are non-void galaxies for comparison (blue dots). The majority of void galaxies are low mass, blue, star forming irregular and spiral galaxies. A red sequence is seen for both the void and non-void galaxy populations, which remains after the colours have been corrected for internal dust reddening. }{\label{colmag}}
\end{figure*}

We present the $k$-corrected $(u-r)$ colour-mass diagram for the GAMA void galaxies as purple stars in the left-hand panel of Fig.~\ref{colmag}. 
The stellar masses and $k$-corrected colours are provided by \citet{2011MNRAS.418.1587T}, with an update to include GAMA~\textsc{ii} galaxies. 
The rest-frame colours are derived from aperture-matched photometry and SED fits of the galaxies optical colours, and are $k$-corrected to $z=0$. See \citet{2011MNRAS.418.1587T} for further details of the colour derivation.
Also plotted for comparison as blue dots are GAMA non-void comparison galaxies to $z\approx0.102$, the highest redshift void galaxy identified in our sample. 
No mass completeness cuts have been applied to the colour-mass diagram, to highlight the fact that a large number of void galaxies are low mass, blue objects with masses $<10^{9}$~M$_{\odot}$: dwarf irregular galaxies that are common in low density environments.  
A red sequence is clearly seen for both the void and non-void galaxies, though the majority of void galaxies are blue systems with masses $<10^{10}$~M$_{\odot}$ (i.e. sub-M$_{\star}$ galaxies). 
At stellar masses $>5\times10^{9}$~M$_{\odot}$, the colour-mass relation is clearly bimodal, with galaxy colours consistent with both star forming and passive galaxies. 
We therefore take $>5\times10^{9}$~M$_{\odot}$ as the mass threshold when comparing the colour and stellar population properties of void and non-void galaxies. 

We also plot colour-mass relation using $(u-r)_{int}$ colours corrected for internal dust reddening in the right-hand panel of Fig.~\ref{colmag}. 
This correction removes galaxies reddened by internal dust from the red sequence, again highlighting the fact that the majority of void galaxies are blue in colour.
In both panels of Fig.~\ref{colmag}, a number of high mass ($>10^{10}$~M$_{\odot}$), red void galaxies with $(u-r) > 1.9$ or  $(u-r)_{int} > 1.6$  hint that at least a fraction of void galaxies have ceased star formation, despite residing in the most under-dense regions of the Universe.  
After the dust reddening is applied, 73 out of 134 of the red sequence galaxies have red colours $(u-r)_{int} > 1.6$, consistent with these galaxies hosting quenched stellar populations.
 We will examine the properties of these red galaxies in greater detail to establish their nature. 
 Given their isolated environment, these massive void galaxies will be used to establish how galaxy mass (as a proxy for halo mass) is responsible for the cessation of star formation. We first examine the colours and spectra of the void galaxies to establish if any have  mission/absorption line features consistent with quenched stellar populations.

\section{Defining a sample of passive galaxies}
\label{sec:red}

\subsection{Active vs. passively evolving void galaxies}
\label{sec:whan}

The red sequence for the general galaxy population contains a mix of star forming, active,  and quenched galaxies \citep[e.g.][]{2010MNRAS.405..783M,2013MNRAS.432..359T,2014MNRAS.437.2521C,2014arXiv1408.5984T}.
To search for signs of ongoing star formation in the red sequence population, we use their GAMA/SDSS spectra to identify  emission lines consistent with ongoing star formation or nuclear activity, and utilise their line strengths to quantify the emission mechanism.  

However, this process is complicated by the fact that the presence of emission lines in a spectrum does not necessarily indicate star formation or nuclear activity. 
\citet{2008MNRAS.391L..29S} show that galaxies identified as LINERs on the Baldwin, Phillips \& Terlevich (1981, BPT) diagram are not necessarily powered by an active nucleus. 
Instead, old, hot, low-mass evolved stars provide enough ionising photons to mimic nuclear activity \citep[e.g.][]{1994A&A...292...13B,2008MNRAS.391L..29S,2012ApJ...747...61Y}. 
Galaxies which appear to host active nuclei through their classification on the BPT diagram are actually \textit{retired} galaxies that have ceased star formation, with old, evolved stars providing their ionising radiation. 
These retired galaxies are actually passive galaxies with spectral features that mimic LINER activity.
For example, using integral field unit spectroscopy, \citet{2013A&A...558A..34B} show that the extended LINER-like emission in NGC~5850 is not confined to its nucleus, but is distributed over the galaxy. 
This emission must therefore be a result of ionisation from distributed sources (likely post-AGB stars), rather than a low-luminosity AGN.

Nuclear activity can also shape a galaxy's stellar population. 
AGN activity is important in the regulation of star formation, preventing hot gas cooling to form stars, and AGN identified through line strength diagnostics are often found on the red sequence or in the green valley. 
We therefore examine the emission/absorption lines of the GAMA void galaxy population to identify those with active stellar populations dominated by emission lines (star formation, AGN, LINERs), vs. those with old stellar populations with featureless absorption line dominated spectra.

The  $W_{\rm H\alpha}$ vs. [N\textsc{ii}]/H$\alpha$ (WHAN) diagram \citep{2011MNRAS.413.1687C} provides a method of separating these different emission mechanisms, including those with spectral lines too weak to be included in the BPT diagram. 
It furthermore allows for low-ionisation nuclear emission-line galaxies to be separated into weakly active AGN, and retired galaxies that have stopped forming stars, with hot, low-mass evolved stars providing their source of ionising radiation. 
\citet{2011MNRAS.413.1687C} find retired and passive galaxies to have near identical stellar populations (indeed, occasionally indistinguishable), having formed no new stars in the past $100$~Myr i.e. they have ceased/retired from star formation. 

\begin{figure*}
\includegraphics[width=0.47\textwidth]{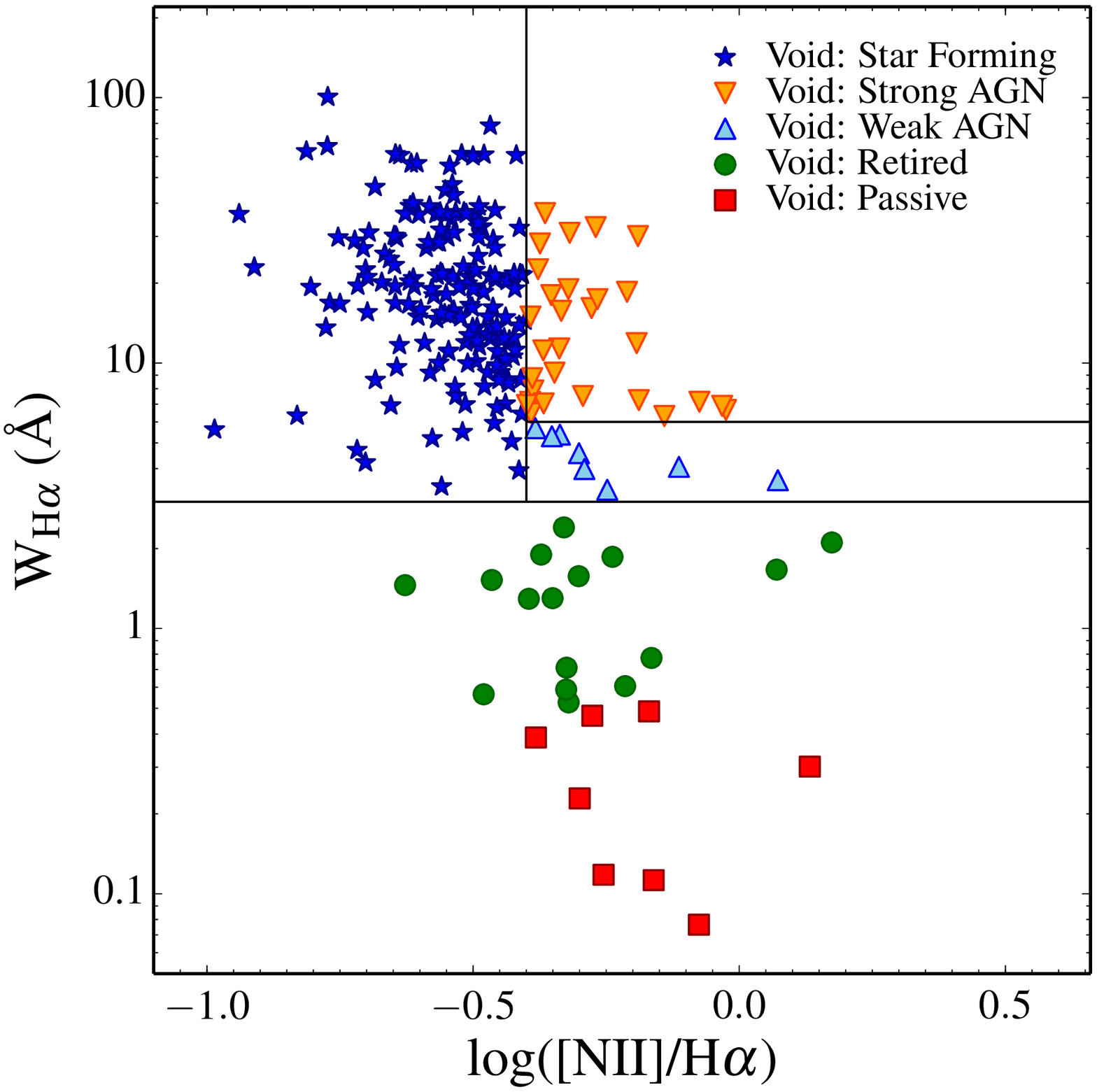}\includegraphics[width=0.47\textwidth]{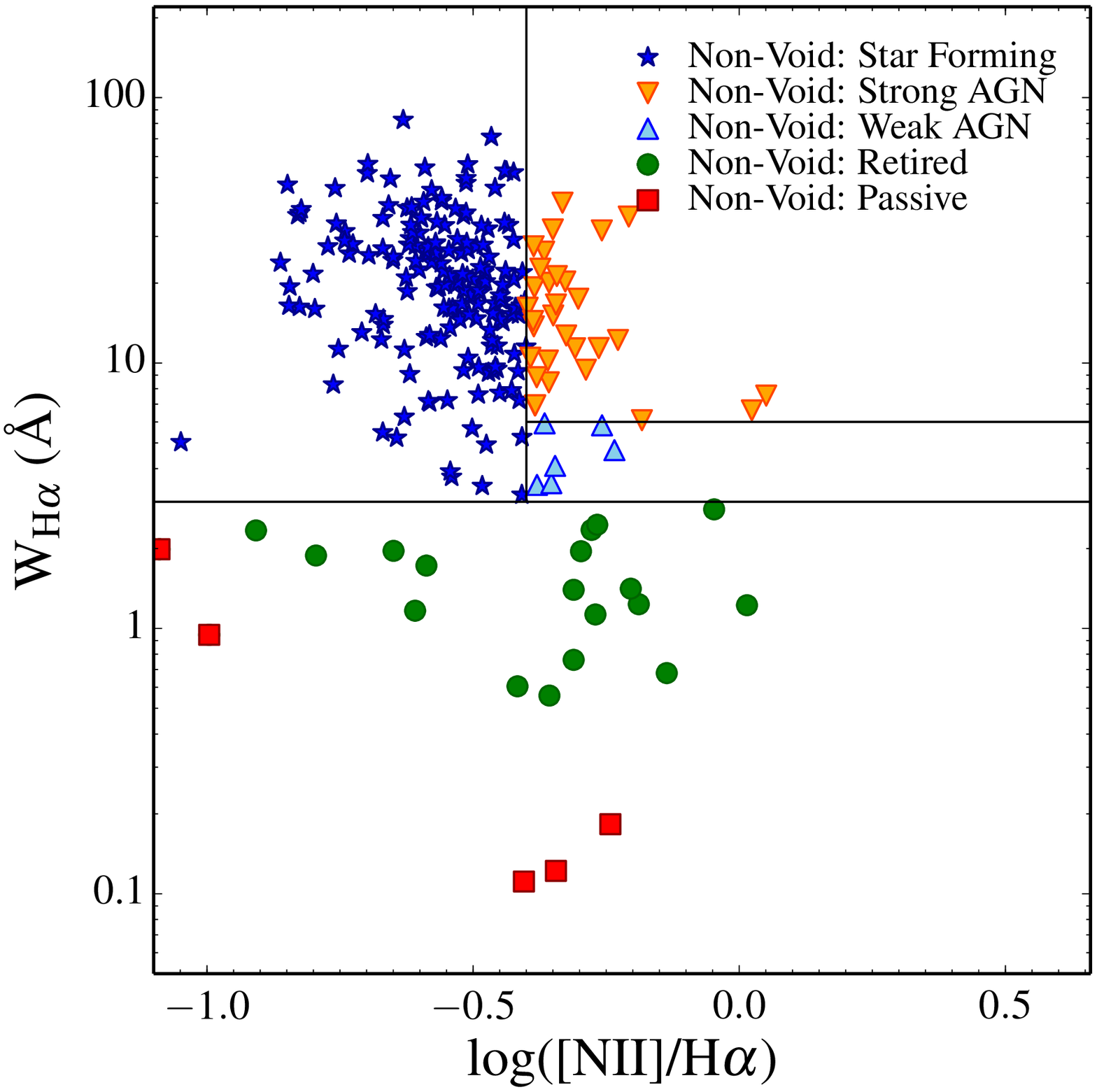}
\caption{\textit{Left-hand panel}: $W_{\rm H\alpha}$ vs. [N\textsc{ii}]/H$\alpha$ (WHAN) diagram \citep{2011MNRAS.413.1687C} for GAMA void galaxies. While the majority of void galaxies are star forming or host AGN, a number of passive/retired galaxies are seen. \textit{Right-hand panel}: the WHAN diagram for a mass-matched randomly-drawn non-void comparison sample.}\label{whan}
\end{figure*}

To separate active from passive/retired galaxies, we construct a WHAN diagram for the GAMA void galaxy population using the strengths of the H$\alpha$ and [N\textsc{ii}]$\lambda6584$, along with the equivalent width of the H$\alpha$ emission line feature. 
For full details of the GAMA line strength catalogue, see \citet{2013MNRAS.433.2764G}. An update is provided for the GAMA \textsc{ii} sample.  
The line widths and fluxes were measured using a flat continuum, rather than stellar population modelling. 
As a result, partially filled stellar absorption lines are provided as absorption, rather than emission, lines, and we discuss such galaxies, along with truly passive galaxies,  in Section~\ref{sec:passabs}
We therefore correct the H$\alpha$ fluxes for stellar absorption following \citet{2003ApJ...599..971H} and \citet{2013MNRAS.433.2764G}. A constant correction of 2.5~\AA\  is applied to the equivalent width of the H$\alpha$ line, and the corrected flux is calculated as follows:

\begin{equation}
F_{\mathrm{H\alpha}} = \frac{\mathrm{EW_{H\alpha} + EW_{C}}}{\mathrm{EW_{H\alpha}}} \times f_{\mathrm{H\alpha}}
\end{equation}

\noindent where $f_{\mathrm{H\alpha}}$ is the observed H$\alpha$ flux, and $\mathrm{EW_{C}}$ is the constant correction factor of 2.5~\AA\ added to the H$\alpha$ equivalent width to correct for stellar absorption.  
A minimum ratio of 3 between  the equivalent width and the error on the equivalent width is required for either the H$\alpha$ or [N\textsc{ii}]$\lambda6584$ line for a galaxy to be included in our line strength analysis diagram in Figure~\ref{whan}.
In future plots, galaxies with this ratio $<3$ have smaller symbols than those with a ratio $>3$.  

We should note here that when examining the optical spectra of void galaxies for signs of activity (either through star formation or nuclear activity), we are examining single fibre spectra. 
This limits us to either the central $2.1$~arcsec of GAMA galaxies with spectra obtained using the AAOmega spectrograph, or the central $3$~arcsec of galaxies with SDSS spectra. 
Galaxies classified as passive via a line strength analysis of their nuclear  spectra may have star forming regions outside of their central bulge. 
If, for example, we are sampling a bulge-dominated spiral galaxy, then we will miss star formation in its spiral arms, as we are sampling the old, passive bulge region of the galaxy with an absence of star formation. 
We will explore this issue in Section~\ref{sec:gamawise}, where we will use mid-IR colours to separate centrally passive from truly passive galaxies. 

The WHAN diagram for GAMA void galaxies more massive than $5\times10^{9}$~M$_{\odot}$ is presented in Fig.~\ref{whan}. Galaxies less massive than this are primarily star forming, so we exclude them from our plot for simplicity. 
The diagnostic lines of  \citet{2011MNRAS.413.1687C} are also plotted. 
Using this diagnostic diagram, galaxies with nuclear spectra consistent with passive or retired stellar populations have H$\alpha$ equivalent widths $0~\mathrm{\AA} <  \mathrm{EW_{H\alpha}} < 3~\mathrm{\AA}$. 
25 galaxies meet this criteria.  
A large population (170) of star forming galaxies is found. 
Eight weak AGN, and 29 strong AGN are also found in the GAMA void sample, though this sample may also contain galaxies with composite AGN and star forming stellar populations.  
Comparable numbers are found for the non-void comparison galaxies.

\subsubsection{Emission-line free galaxies}
\label{sec:passabs}

The WHAN diagram (and, indeed, any line strength diagnostic diagram), does not include galaxies with H$\alpha$ in absorption, and the GAMA line strength classification does not fit stellar population models prior to fitting a galaxy. 
As such, we miss galaxies with partially infilled H$\alpha$ absorption, and those dominated purely by absorption lines in the WHAN diagram. 

In our sample, sixteen void galaxies have equivalent widths measured for their [N\textsc{ii}]$\lambda6584$ line, but do not have any measured H$\alpha$ emission. 
Indeed, a visual inspection of their spectra reveals a number to have slight infill of the H$\alpha$ lines due to emission.
Eleven have [N\textsc{ii}]$\lambda6584$ equivalent widths $>0.5$~\AA\, and we therefore class them as retired galaxies in future plots as a heating source is required to produce this emission. 
 Any remaining galaxies without H$\alpha$ or [N\textsc{ii}]$\lambda6584$ in emission are classified as passive in future plots, and five galaxies meet these criteria. 

This classification scheme results in 40 galaxies more massive than $5\times10^{9}$~M$_{\sun}$ without a line strength classification. 
For such galaxies, the emission line fitting code was not able to accurately measure either their [N\textsc{ii}]$\lambda6584$ or H$\alpha$ lines, and hence their stellar population could not be characterised. 
These galaxies span the complete mass range, but the majority have colours $(u-r) > 1.9$, where the galaxy population is dominated by passive/retired galaxies or edge-on, dust-reddened disc galaxies. We plot such objects in future plots as grey hexagons.

\subsection{The colour-mass relation for galaxies with WHAN classifications}
\label{sec:whancol}

\begin{figure*}
\includegraphics[width=0.47\textwidth]{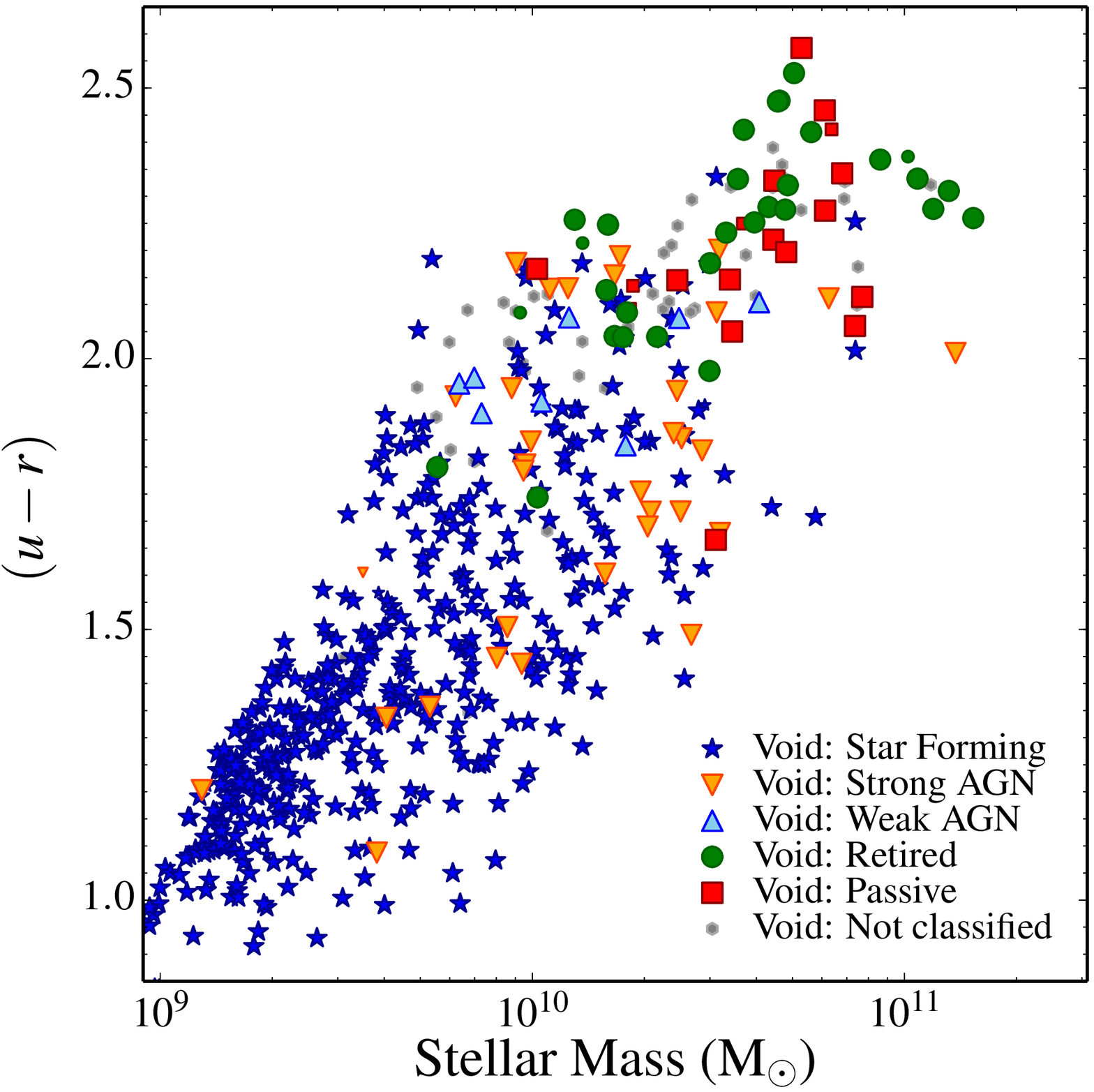}\includegraphics[width=0.47\textwidth]{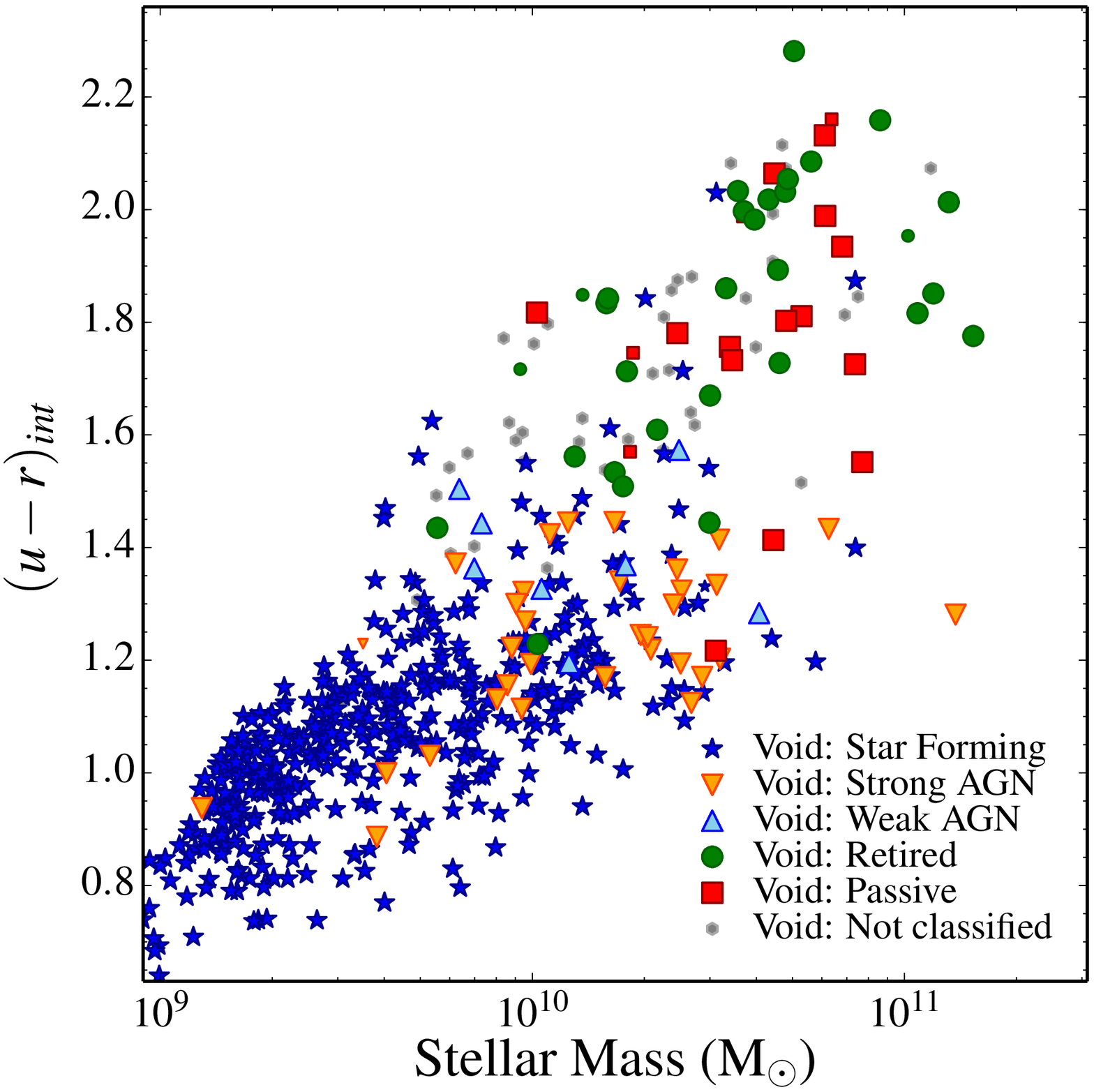}
\includegraphics[width=0.47\textwidth]{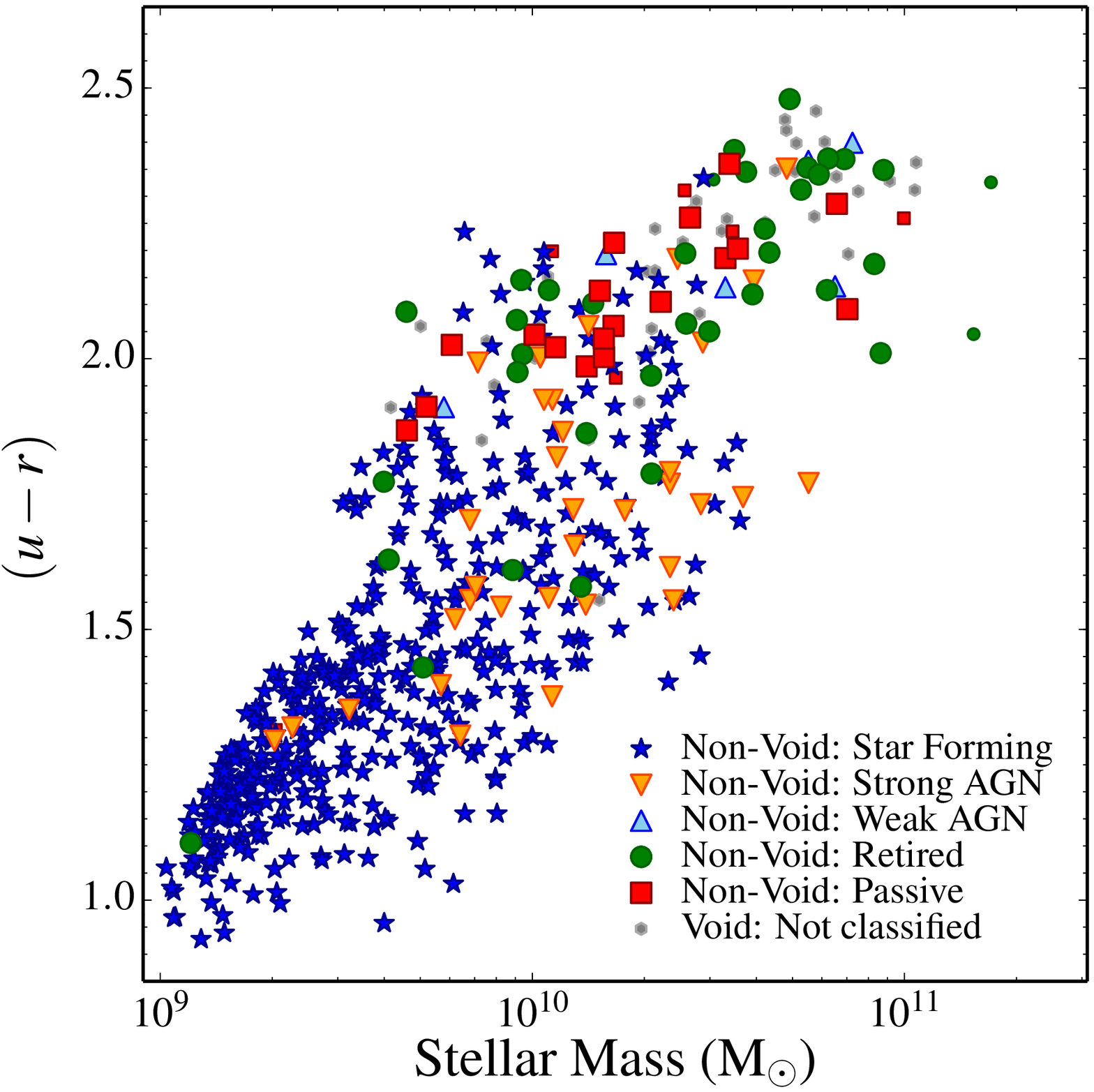}\includegraphics[width=0.47\textwidth]{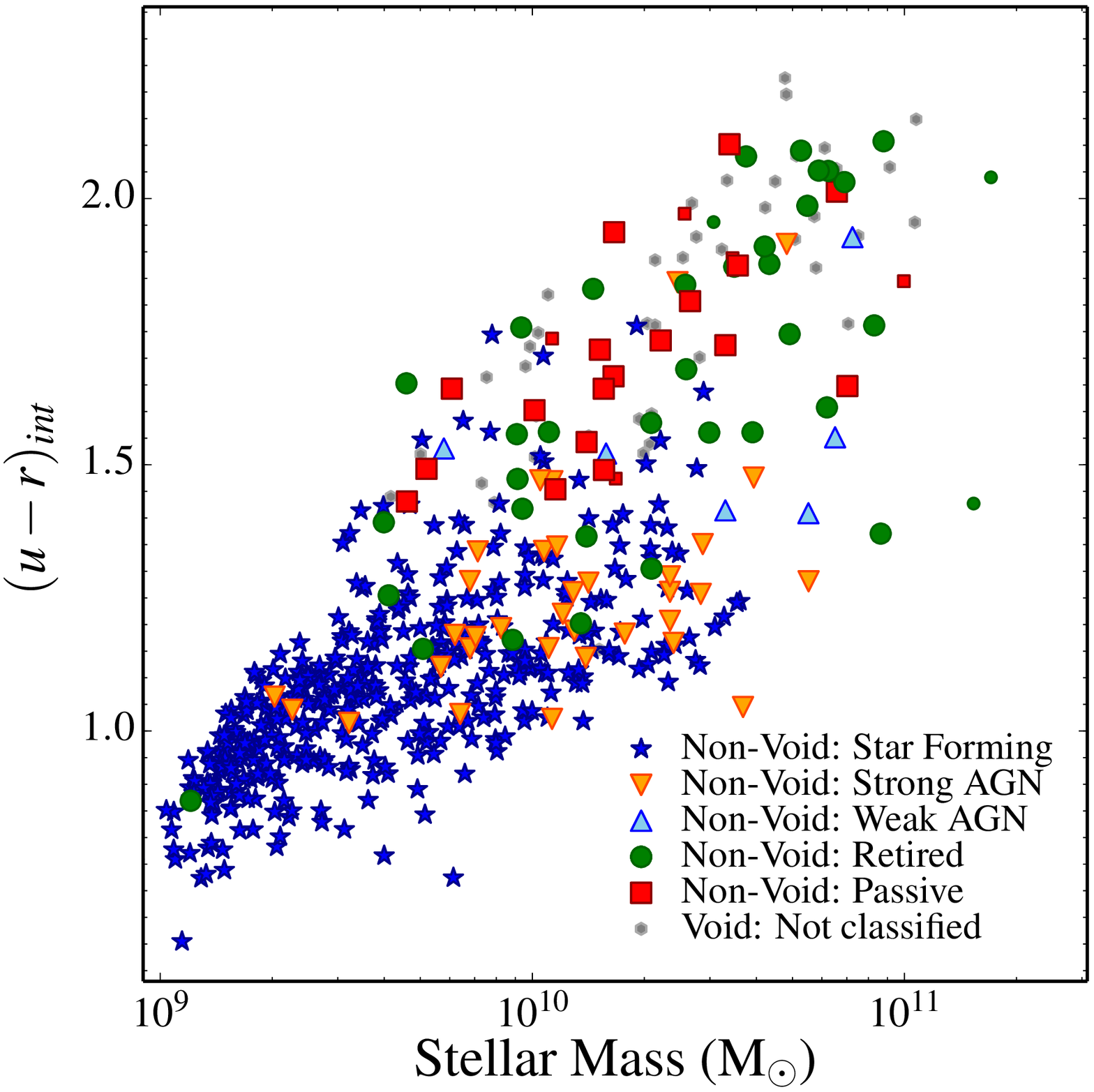}
\caption{The $(u-r)$ vs. stellar mass diagram for GAMA void galaxies (top row), and for a matched random sample of non-void comparison galaxies (bottom row). The symbols have the same meaning as in Fig.~\ref{whan}. The smaller versions of the symbols have an equivalent width vs. equivalent width error ratio $<3$. After the correction for internal dust reddening, star forming galaxies and galaxies with strong AGN-like features move off the red sequence, and only passive/retired galaxies remain.}\label{cmdwhan}
\end{figure*}

Do the colours of the void galaxies reflect their line strength measurements? To answer this, in Figure~\ref{cmdwhan} we reconstruct the $(u-r)$ colour-mass diagram for void galaxies based on their classification on the WHAN diagram. Galaxies with H$\alpha$ in absorption that could not be included in the WHAN diagram are added as passive/retired galaxies, depending on the strength of their [N\textsc{ii}]$\lambda$6583 line. 
The top left-hand panel shows their observed $(u-r)$ colours, and the $(u-r)$ colours presented in the top right-hand panel have been corrected for internal dust reddening. 
The colours have been $k$-corrected to $z=0$ in the plots, and corrected for foreground extinction. 
When corrected for intrinsic dust reddening, strong AGN are almost completely removed from the red sequence, with only red and dead void galaxies having retired or passive stellar populations remaining. 

As a comparison, we include a sample of randomly drawn, matched non-void galaxies. 
For each void galaxy, we select a non-void comparison galaxy of similar mass ($\pm 20$~per~cent~M$_{\star}$), rest-frame, colour uncorrected for dust reddening ($(u-r) \pm 0.15$), and redshift ($z\pm0.01$) to ensure we are selecting galaxies at a comparable stage in their evolution. 
Few high-mass galaxies with AGN-like line ratios are seen in both samples, with the red sequence dominated by passive and retired galaxies at $M_{\star} > 3\times10^{10}$~M$_{\odot}$. 

The distribution of galaxies by WHAN classification on the colour-mass diagram is remarkably similar for the void and comparison galaxies. 
A K-S test reveals that for both the void and randomly-drawn mass-matched comparison sample, the colour distributions of the two populations are identical.
 The p-values are $>0.1$ that both the void and non-void samples are drawn from the same population for all five WHAN classes. 
When we repeat this K-S test using galaxy mass rather than colour, both the void and the field samples are drawn from the same population for four of the WHAN classes, with the exception of the passive galaxies
The passive galaxies have $p=0.034$ that the void and non-void samples have the same mass distribution. 
While we cannot completely rule out that passive void and non-void galaxies are drawn from the sample mass distribution, it can be seen from Fig.~\ref{cmdwhan} that the passive comparison galaxies extend to lower galaxy mass than the void population.

\subsection{Colour cuts to define passively evolving galaxies}
\label{sec:colcuts}

Using a combination of WHAN line strength diagnostics and optical $(u-r)$ colours, we show in Fig.~\ref{cmdwhan} that the void and non-void red sequence is dominated by galaxies with old stellar populations. 
We can therefore use a combination of optical colours and line-strength diagnostics to define a colour cut above which the majority of galaxies are non star forming. 
We use a lower mass limit $M_{\star} > 5\times10^{9}$~M$_{\odot}$ when selecting passive galaxies- below this limit, blue galaxies dominate both the void and non-void populations, and there is no evidence that isolated galaxies below this limit are quenched \citep{2012ApJ...757...85G}. 
Red galaxies below this mass are typically dwarf satellites, which are strongly affected by environment  and must be  excluded in a study of mass quenching.

\begin{figure}
\includegraphics[width=0.47\textwidth]{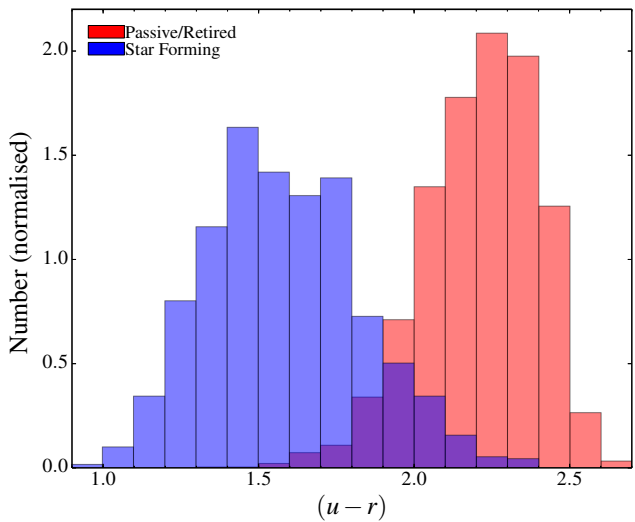}
\caption{Histogram showing the fraction of star forming and passive galaxies as a function of $(u-r)$ colour for the non-void galaxy population. The galaxies were selected via their WHAN diagnostic into star forming or passive. The passive sample includes retired galaxies with hot, old, low mass stars as their ionisation source. We exclude AGN from this plot for simplicity. The fraction of star forming galaxies drops as $(u-r)$ becomes redder.}\label{cols}
\end{figure}

We identify a $(u-r)$ colour cut consistent with a passively evolving stellar population from the comparison sample defined in Section~\ref{sec:fieldsamp}. 
The non-void comparison sample is used due to its larger sample size vs. the void galaxy sample. 
First, we identify passive vs. star forming galaxies using line strength diagnostics.  
All comparison galaxies more massive than $5\times10^{10}$~M$_{\sun}$ are classified according to their location on the  WHAN line strength diagnostic diagram. 
This analysis is described in full in Sec.~\ref{sec:whan}. 
Galaxies with line strengths that quantify them as AGN are removed from the sample.  
This leaves two remaining classes of galaxy: star forming galaxies dominated by a young stellar population, and passive/retired galaxies whose optical spectra are dominated by old stars. 
We then construct a histogram showing the $(u-r)$ colour distributions of the two samples (Fig.~\ref{cols}).
It can be seen from Fig.~\ref{cols} that for colours $(u-r)>1.9$, passive/retired galaxies dominate in terms of number.
This distribution in colour by separating star forming vs. passive galaxies is similar to the bimodal colour distribution of lower luminosity galaxies in \citet{2012MNRAS.426.3041H}. Low luminosity galaxies dominate both samples in terms of number.
We therefore take $(u-r)=1.9$ as the lower limit of galaxy colour when defining a sample of red void galaxies with optical colours consistent with a passively evolving stellar population.

Selecting passive galaxies based on colour alone is not perfect, and it can been seen from Fig.~\ref{cols} that passive and star forming galaxies overlap in colour when $(u-r) > 1.5$. 
This overlap of passive and star forming galaxies remains after a model-dependent correction for intrinsic dust reddening has been applied, and we therefore choose to use uncorrected $(u-r)$ colours when selecting our red comparison and void samples. 
This separation may become more complicated when examining void vs. non-void galaxies. \citet{2012MNRAS.426.3041H} show that at a given luminosity, void galaxies are typically bluer than their non-void counterparts by $(u-r)\approx0.1$. 
However, our cut ensures that we will primarily be comparing galaxies that are non-star forming in both the void and non-void samples.

\subsection{Obscured star formation}

Searching for star formation at optical wavelengths has drawbacks. 
Dust obscuration can mask low levels of star formation, such that the object's spectrum will appear passive. 
To examine if obscured star formation is present in the void red-sequence galaxy population, we therefore go on to examine their mid-IR properties. 
The mid-IR is a more sensitive tracer of recent star formation than the optical. PAH emission is a typical feature of star forming galaxies. 
Longer than 8~$\mu$m, emission from dust heated by younger stars begins to trace star formation.  The optical colours of the galaxies will redden and they will join the red sequence within 1-2~Gyr of the cessation of star formation. 
The heating of circumstellar envelopes is the main source of mid-IR emission in galaxies, and this is sensitive to star formation over relatively long timescales ($>1$~Gyr).

The optical red sequence contains not only genuinely passive, non-star forming galaxies, but a number of late-type galaxies with their colours reddened by dust extinction or with a low level of star formation which is not sufficient to move them to the blue cloud.  
This low level of star formation would not be picked up in the fibre spectroscopy we examine in this work.  
We instead use WISE mid-IR colours as a tracer of recent star formation in galaxies classed as passive/retired based on colour and line-strength diagnostics in Section~\ref{sec:colcuts}.

\section{GAMA-WISE}
\label{sec:gamawise}

The \textit{Wide-field Infrared Survey Explorer (WISE)} telescope provides this mid-IR data for the GAMA survey. 
The 3.4~$\mu{\rm m}$ (W1) and 4.6~$\mu{\rm m}$ (W2) \textit{WISE} bands trace the continuum light from evolved stars. 
W1 is most sensitive to stellar light, and W2 is also sensitive to hot dust. 
W1$-$W2 is therefore a good colour for identifying galaxies dominated by AGN emission \citep{2011ApJ...735..112J}. 
The 12~$\mu{\rm m}$ W3 band traces the 9.7~$\mu{\rm m}$ silicate absorption feature, as well as 11.3~$\mu{\rm m}$ PAH and NeII emission line. 
The W4 band traces the warm dust continuum at 22~$\mu{\rm m}$, and is used to trace AGN activity and reprocessed radiation from star formation. 
Thus by examining the colours of galaxies in the mid-IR, we can compare the recent star formation histories of void vs. non-void galaxies \citep[e.g.][]{2013AJ....145....6J}. 

In particular, we are interested in revealing star formation in red galaxies via the W3 flux (12.0~$\mu{\rm m}$) which will highlight void galaxies with current star formation. 
\citet{2012ApJ...748...80D} find that 80~per~cent of the 12~$\mu$m emission in star forming galaxies is produced by stellar populations younger than 0.6~Gyr,  
\textit{WISE} is therefore ideal for identifying galaxies with low levels of nuclear activity and star formation that are not easily found in their optical spectra.

\subsection{\textit{WISE} photometry}

GAMA sources with \textit{WISE} photometry are identified by cross-matching the \textit{WISE} All-Sky Catalogue to the GAMA II observed sources catalogue using a $3$~arcsec cone search radius. 
86~per~cent of GAMA sources in G09 were detected in the WISE All-Sky survey, with 82~per~cent of G12 sources and 89~per~cent of G15 sources detected. 
The data product and its reduction is described in full detail in \citet{2014ApJ...782...90C}. 
Here, we match the optically selected void galaxy sample defined in Section~\ref{sec:vgals} to the GAMA-\textit{WISE} catalogue using the galaxies GAMA catalogue IDs. 
527 out of 577 void galaxies with masses $>10^{9}$~M$_{\odot}$ are matched to the GAMA-\textit{WISE} photometry catalogue. For the comparison sample, 7233 out of 7718 matches are found.

The majority (70~per~cent) of non-matches in both the void and comparison samples have masses $<5\times10^{9}$~M$_{\odot}$- 10~per~cent of all galaxies below this mass limit.  
We therefore exclude all galaxies less massive than this mass due to incompleteness, leaving 295 of the \textit{WISE}-void galaxy matches  and 4978 non-void galaxies, more massive than $5\times10^{9}$~M$_{\odot}$ in our \textit{WISE} sample.  
 
For the void sample, 10 non-matches with masses $>5\times10^{9}$~M$_{\odot}$ vs. 295 matched galaxies are found, approximately 3~per~cent of the sample above this mass limit. All are less massive than $1.6\times10^{10}$~M$_{\odot}$ and have $(u-r) > 1.6$. 
The non-matches in the comparison sample with masses $>5\times10^{9}$~M$_{\odot}$ exhibit a wider range in mass and colour than the void non-matches. 
Similar to the void galaxy sample, these non-matches are a very small fraction of the comparison galaxies ($<3$~per~cent), and we therefore do not expect the absence of the non-matches to have a significant impact on our results.

Fewer than $20$~per~cent of targets in the GAMA equatorial fields have S/N$>2$ in the W4 band, and we therefore exclude the W4 data from our examination of void galaxy colours, instead utilising W3 as a tracer of star formation. 
For both samples, 30~per~cent have W3 magnitudes (and therefore $[4.6]-[12]$ colours) that are flagged as upper limits or null photometry.
Where available, upper limits on their W3 photometry from the profile-fitted measurements from the WISE All-Sky
Catalogue are provided.
We include these upper limits in our examination of the WISE colour-colour diagram in Figures~ \ref{ircolcol}, \ref{wisewhan}, and \ref{irmass} as upper limit arrows. 

In this paper, we present \textit{WISE} colours in Vega magnitudes, the native magnitude system of the \textit{WISE} dataset. The Vega magnitudes can be transformed to the AB-System using the transformations of \citet{2011ApJ...735..112J}. 
However, care must be taken when applying these transformations due to uncertainty in the W3 and W4 filter response curves (see \citealt{2010AJ....140.1868W}, and \citealt{2014PASA...31...49B} for more information).
We therefore choose to work in the Vega magnitude system in order to compare the mid-IR colours of the GAMA galaxies to the literature. 
The Vega magnitudes are then $k$-corrected to $z = 0$ using the spectral energy distributions of \citet{2014ApJS..212...18B}.

\subsection{Optically red void galaxies in the mid-IR}

We investigate if the void and non-void comparison galaxies with $(u-r)$ colours consistent with quenched stellar populations are truly passive via their mid-IR colours. 
To do this, we utilise the red sequence galaxy samples defined in Section~\ref{sec:colcuts}, with stellar masses $>5\times10^{9}$~M$_{\odot}$ and colours $(u-r)>1.9$ consistent with a non star forming stellar population.
The red galaxies identified in Section~\ref{sec:colcuts} are matched to the GAMA \textit{WISE}  photometry catalogue, with 95 void and 2125 non-void galaxies with line strength measurements having mid-IR photometry (31~per~cent and 50~per~cent of the WISE void and comparison samples respectively).

\begin{figure}
\includegraphics[width=0.47\textwidth]{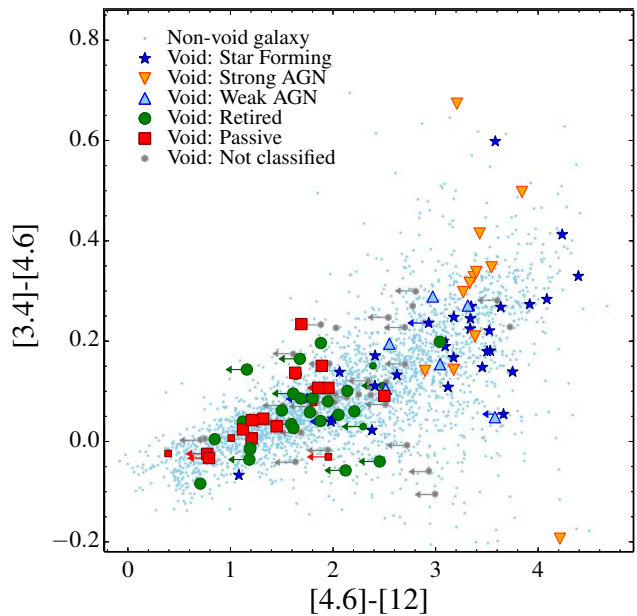}
\caption{\textit{WISE} $[4.6]-[12]$ vs. $[3.4]-[4.6]$ ([W2--W3] vs. [W1--W2]) colour-colour diagram for void and non-void comparison galaxies with masses $>5\times10^{9}$~M$_{\odot}$ and passive optical colours $(u-r)>1.9$.  The smaller versions of the symbols have an equivalent width vs. equivalent width error ratio $<3$. The non-void galaxy population occupying the same region of the CMD is also plotted for comparison (blue dots). A range of $[4.6]-[12]$ colours is seen for both the void and comparison galaxy populations, consistent with a range of recent star formation activity. }\label{ircolcol}
\end{figure}

The $[3.4] - [4.6]$ vs. $[4.6]-[12.0]$  colour-colour diagram for the GAMA red sequence void and comparison samples are shown in Figure~\ref{ircolcol}. 
The symbols of the void galaxies in Fig.~\ref{ircolcol} reflect their WHAN diagnostic.
Also plotted for comparison is the non-void sample as blue dots. 
Most obvious in this plot is how few of the optically red void galaxies with $(u-r) > 1.9$ have $[4.6]-[12.0]$ colour $< 1.5$. 
Galaxies with $[4.6]-[12.0]<1.5$  are typically quenched galaxies such as ellipticals with negligible star formation, though the red sequence void galaxies exhibit a wide range of mid-IR colour.   
The range of  $[4.6]-[12.0]$ colours exhibited is large for both the void and non-void central populations, with $ 0 \lesssim [4.6]-[12.0] \lesssim 4.4$. 
This spread in colour reflects different levels of ongoing star formation activity for the two populations, from quenched galaxies and giant ellipticals ($[4.6]-[12.0] < 1.5$) through to starburst galaxies with $[4.6]-[12.0]$ colours  $> 3.0$. 
Several optically red void galaxies with AGN-like spectral features are seen with blue $[4.6]-[12.0]$ colours, and these are likely composite objects with active nuclei and ongoing star formation. 

Using their $[4.6]-[12.0]$ colours as a proxy for ongoing/recent star formation, we compare the distribution of $[4.6]-[12.0]$ colour for the red void and non-void populations to search for differences between void galaxies and the general galaxy population. 
For this comparison, we only utilise non-void comparison galaxies in the same mass range as the void galaxies, with masses $ 5\times10^{9}~{\rm{M_{\sun}}}< M_{\star} < 1.5\times10^{11}~{\rm{M_{\sun}}}$.
A two-sided K-S test gives a p-value$=0.42$, and we therefore reject the hypothesis that the void and non-void galaxy samples are clearly drawn from different populations. 
An identical result is found for the $[3.4]-[4.6]$ colours of the void and non-void galaxies, with a p$=0.22$. Again, that the two populations appear to be drawn from the same colour distribution.

\subsection{WHAN-classified galaxies in the mid-IR}
\label{sec:wisewhan}

\begin{figure*}
\includegraphics[width=0.47\textwidth]{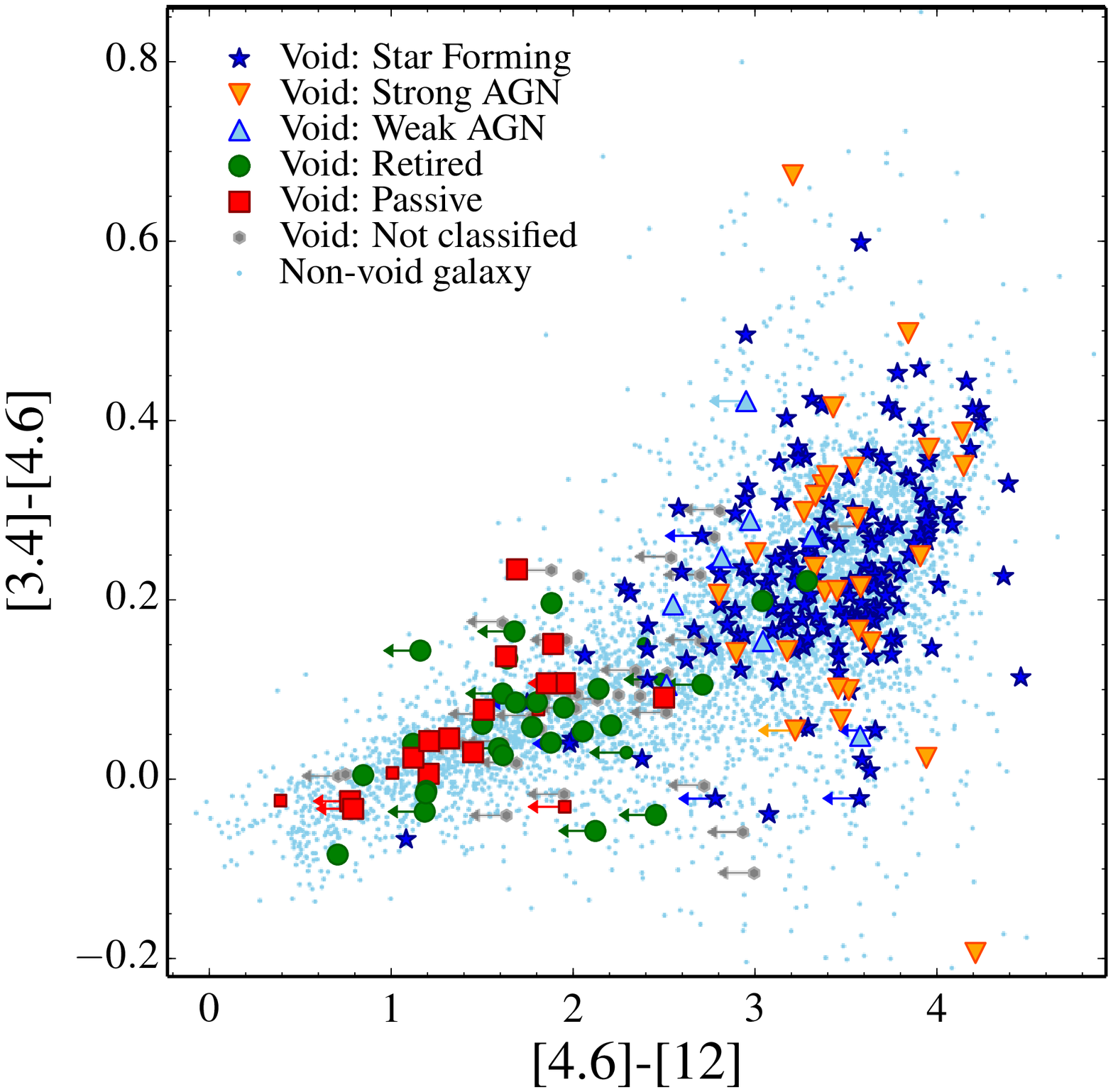}\includegraphics[width=0.47\textwidth]{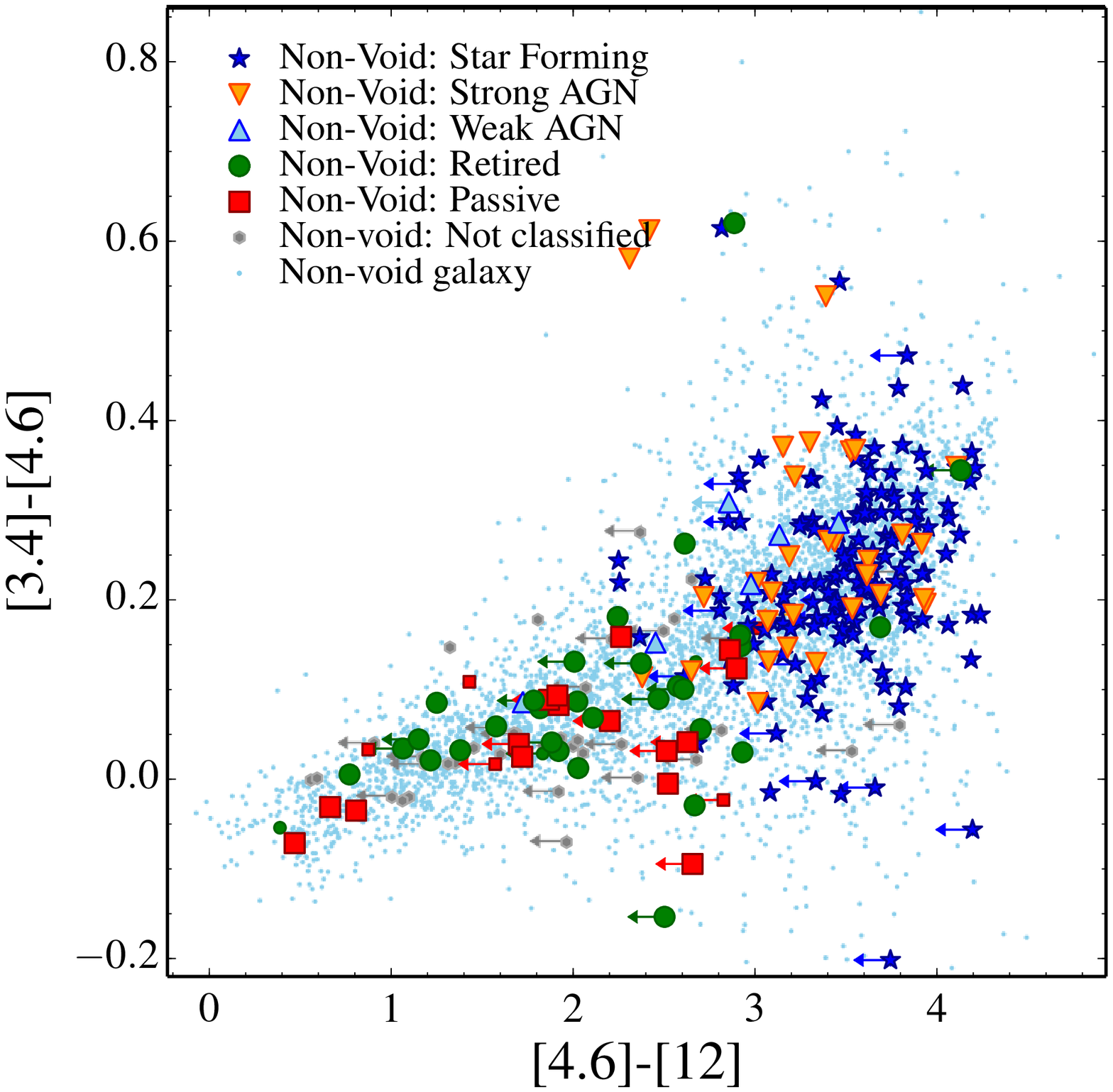}
\caption{\textit{Left-hand panel}: \textit{WISE} $[3.4]-[4.6]$ vs. $[4.6]-[12]$ colour-colour diagram for the GAMA void galaxy sample. The colours of the points correspond to their classification on the WHAN diagram presented in Fig.~\ref{whan}. The majority of void galaxies with $[4.6]-[12] < 2.5$ have optical nuclear spectra consistent with retired/passive stellar populations. However, a number of these galaxies likely host obscured star formation or are forming stars outside of their nuclei, with $[4.6]-[12]>1.5$ colours inconsistent with a truly passive stellar population. \textit{Right-hand panel}:  \textit{WISE} $[3.4]-[4.6]$ vs. $[4.6]-[12]$ colour-colour diagram for a mass-matched randomly-drawn sample of non-void galaxies.}\label{wisewhan}
\end{figure*}

The mid-IR bands covered by \textit{WISE} are particularly sensitive to obscured star formation, or enhanced ISM emission (e.g. from nuclear activity), and we might therefore be able to use a combination of mid-IR colour and WHAN classification to better understand the true nature of the stellar populations in these galaxies. 
Using the emission-line classifications presented in Section~\ref{sec:whan}, we examine the mid-IR colours of the galaxies with the nature of their stellar activity classified using the WHAN diagram. 
The $[4.6]-[12]$ vs. $[3.4]-[4.6]$ colour-colour diagram for the GAMA void galaxy sample is replotted in the left-hand panel of Fig.~\ref{wisewhan} for galaxies classified according to their line strength ratios on the WHAN diagram. 

Of the 261 void galaxies with WHAN classifications identified in Section~\ref{sec:whan} with masses $>5\times10^{9}$~M$_{\odot}$, 254 are present in the GAMA-\textit{WISE} catalogue. 
From these, 30 have $[4.6]-[12]$ colours derived from upper limits on their photometry.  
 For the non-void comparison sample, 255 have GAMA-\textit{WISE} matches, of which 43 have $[4.6]-[12]$ colours derived from upper limits on their W3 photometry. 

As Fig.~\ref{wisewhan} illustrates, both the void and non-void samples show trends in $[4.6]-[12.0]$ colour dependent on their  WHAN classification. 
Passive/retired galaxies tend towards bluer mid-IR colours, whereas star forming and AGN classified galaxies have redder mid-IR colours. 
It is clear, however, that passive/retired galaxies in both the void and comparison sample do not have mid-IR colours that reflect their line strength classifications. 
Galaxies with $[4.6]-[12] < 1.5$ are consistent with no ongoing star formation \citep{2011ApJ...735..112J,2014ApJ...782...90C}, and we utilise this cut to separate passive vs. star forming galaxies.  
The locus of elliptical galaxies is located at $[4.6]-[12] \approx 0.5$, with spiral/starburst galaxies having $[4.6]-[12] > 2$. 
 Fig.~\ref{wisewhan} shows that the passive/retired galaxies for both the void and comparison samples have $0 < [4.6]-[12] < 4$, spanning a wide range of current star formation activity.  
As we are using nuclear spectra in our emission line diagnostics when selecting these passive/retired galaxies, we may miss light from non-central star forming regions in these galaxies, else their star formation is obscured in the optical.
Galaxies classed as star forming via WHAN diagnostics in both the void and comparison samples have $[4.6]-[12] > 1.5$, and lie on the region of the $[3.4]-[4.6]$ vs $[4.6]-[12]$ \textit{WISE} colour-colour diagram occupied by spiral/star forming galaxies. 
The WHAN classifications for star forming galaxies therefore accurately reflect their current state of star formation activity in the mid-IR.

Galaxies classed as AGN using WHAN line strength diagnostics have $[4.6]-[12] > 2.5$, and the colours of these galaxies overlap with star forming galaxies. 
\citet{2012ApJ...753...30S}  take $[3.4]-[4.6] > 0.8$ as the mid-IR colour criterion for luminous, X-ray selected AGN.
None of these galaxies in either the void or comparison samples meet this criterion. They do, however, meet the AGN criteria in \citet{2011ApJ...735..112J}, where AGN can have $[3.4]-[4.6] $ as low as  0.5. Meanwhile, weak seyferts can have colours lower than this threshold. 
These galaxies have  values of $[4.6]-[12]$ consistent with them being actively star forming or hosting AGN activity, but blue $[3.4]-[4.6] \approx 0$ inconsistent with them hosting powerful AGN, though they may contain weak seyferts. 
The objects are low luminosity AGN or LINERs with high levels of star formation i.e. they are composite AGN, producing mid-IR SEDs that overlap with non-AGN star forming galaxies.

\subsection{Galaxy Mass and the cessation of star formation}
\label{sec:irmassquench}

Massive galaxies in all environments are typically red, with quenched star formation. 
By investigating how galaxy colour changes as a function of stellar mass (as a proxy for halo mass in isolated galaxies), we can examine the role mass plays in the quenching of star formation.  
As optical colours and line-strength diagnostics alone are not sufficient to identify truly passive galaxies, as we show in Section~\ref{sec:wisewhan}, we utilise the GAMA \textit{WISE} catalogue to examine this effect. 
At mid-IR wavelengths, the polycyclic aromatic hydrocarbon feature at 11.3~$\mu$m is excited by UV radiation from young stars, and  \textit{WISE} $[4.6]-[12]$ colours are a more sensitive diagnostic of recent star formation than optical colours. 
While the warm dust continuum at $\sim22$~$\mu$m is a better tracer of star formation, few void galaxies have sufficient S/N for detections in this band. 
We utilise this colour diagnostic to search for mass quenching in the GAMA void galaxy population, using the cut $[4.6]-[12]= 1.5$ from \citet{2014ApJ...782...90C}  to separate passive objects from galaxies dominated by star formation. 
See \citet{2014ApJ...782...90C} for a more thorough discussion of using the \textit{WISE} 12~$\mu$m band and  $[4.6]-[12]$ colour for star formation diagnostics.

\begin{figure*}
\begin{center}
\includegraphics[width=0.98\textwidth]{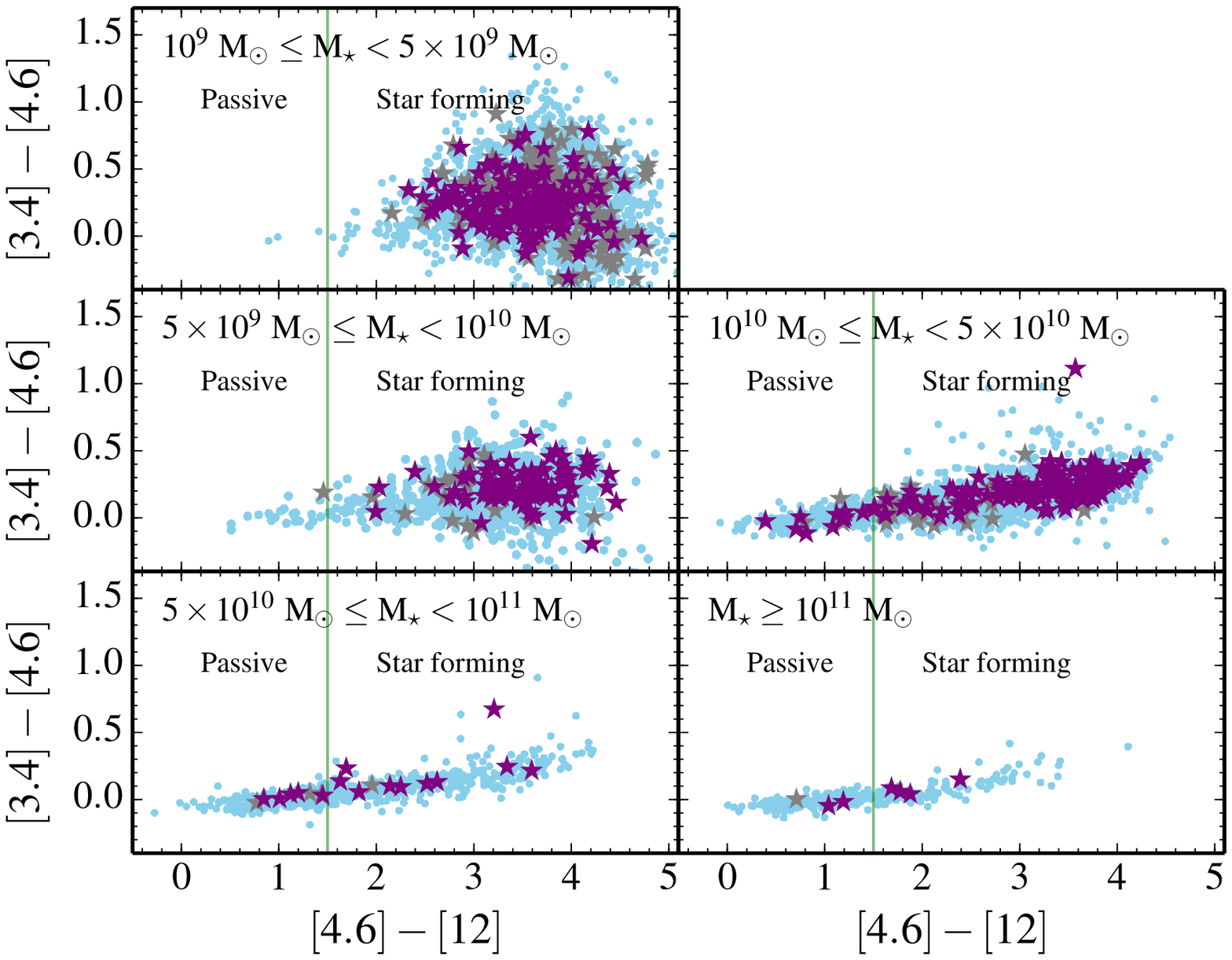}
\end{center}
\caption{\textit{WISE} $[3.4]-[4.6]$ vs. $[4.6]-[12]$ colour-colour diagram for void and non-void comparison galaxies (purple stars and blue dots respectively). Void galaxies with upper limits on their W3 photometry are shown as grey stars.  Five mass bins are shown, as labelled in the top left-hand corner of each panel. The vertical green line is at $[3.4]-[4.6] = 1.5$, the divide between passive vs. star forming galaxies. The comparison and void samples overlap at all masses.  Passive void galaxies bluer than $[4.6]-[12] = 1.5$ have masses M$_{\star} > 10^{10}$~M$_{\odot}$, showing that in the most low density regions of the Universe, a fraction of massive galaxies have ceased star formation. Unlike their non-void counterparts, massive void galaxies have not undergone extreme environmental quenching, and will be star forming unless they reside in sufficiently massive haloes for intrinsic quenching processes to become efficient. }\label{irmass}
\end{figure*}

To search for a relation between galaxy stellar mass and star formation activity for the void galaxy population, we examine their $[3.4]-[4.6]$ vs. $[4.6]-[12]$ colours binned by mass in Fig.~\ref{irmass}. 
The galaxies are split into five mass bins, from  $M_{\star} >10^{9}$~M$_{\odot}$ to $M_{\star} < 5\times10^{11}$~M$_{\odot}$. 
The void galaxies are plotted as purple stars, with upper limits on the void galaxy colours are plotted as grey stars.
Also plotted for comparison is the non-void sample as blue dots, again binned by mass.  
We utilise a colour cut  $[4.6]-[12] < 1.5$ to identify void galaxies with passive mid-IR colours.

The colour-colour diagram changes with galaxy mass. 
For the lowest mass bin ($<5\times10^{9}$~M$_{\odot}$), both the void and non-void galaxies predominantly have IR colours consistent with star formation, with $[4.6]-[12] > 1.5$. 
 It is this region of the \textit{WISE} colour-colour diagram in which optically selected blue void galaxies with H$\alpha$ emission indicative of star formation reside (Figure~\ref{wisewhan}). 
There is also a spread in the $[3.4]-[4.6]$ colours of objects in this low mass bin, showing a spread in the level of hot dust emission in these star forming galaxies.  
This spread in $[3.4]-[4.6]$ colours tightens with increasing galaxy mass. 

As the mass of the galaxies increases, the $[4.6]-[12]$ colours become increasingly dominated by passive colours, with this trend seen for both the void and non-void populations. 
For galaxies more massive than $10^{11}$~M$_{\odot}$, starburst and spiral galaxies with $[4.6]-[12] > 3.0$ are almost entirely absent, and the colours of these galaxies instead overlap with those of elliptical galaxies and spirals with low levels of star formation. 
Regardless of environment, more massive galaxies exhibit lower levels of star formation.

Twenty-six void galaxies have passive mid-IR colours, and no passive void galaxies with reliable photometry exist in our sample with masses $<10^{10}$~M$_{\odot}$. 
These passive, high mass void galaxies do not exhibit tidal tails or shells in SDSS imaging (they have not undergone recent mergers), and do not exhibit edge-on disc morphologies. 
Furthermore, these massive void galaxies all have local galaxy surface density $\Sigma_{5} < 0.35$~Mpc$^{-2}$, which \citet{2013MNRAS.435.2903B} class as being a low-density environment. 
They have 5th nearest neighbour distances $>5$~Mpc, and are well isolated in SDSS colour imaging, and are therefore unlikely to be cluster interlopers.
Galaxies with stellar masses $>10^{10}$~M$_{\odot}$ typically occupy haloes with masses $>10^{12}$~M$_{\odot}$, where various processes such as AGN heating and virial shock heating may prevent gas cooling into new stars, quenching star formation \citep[e.g.][]{2006MNRAS.368....2D,2006MNRAS.365...11C}.

\section{The highest mass void galaxies}
\label{sec:higmass}

The twenty-six passive void galaxies identified in Section~\ref{sec:irmassquench} have masses $>10^{10}$~M$_{\odot}$, where the mass assembly histories of galaxies are increasingly dominated by major mergers \citep{2003MNRAS.341...54K,2004ApJ...600..681B}. 
These major mergers between galaxies of similar mass ($\approx3:1$) will transform the morphology of a galaxy from disc to bulge dominated, removing structures such as spiral arms and bars. 
Alternatively, given their isolation, these isolated galaxies may have increased in mass due to star formation, becoming quenched above some critical mass threshold, and would retain their discy morphologies.
We examine SDSS colour imaging for the 48 most massive GAMA void galaxies (regardless of their colour) to search for these features, with this high-mass sample presented in Fig.~\ref{massive}. 
These massive void galaxies have colours $1.70 <(u-r) < 2.57$, though only two have $(u-r) < 1.9$. 
The majority have optically passive colours using the colour cut defined in Section~\ref{sec:colcuts}.
The galaxies are sorted from left to right in order of increasing mass, with masses $3.6 \times10^{10}~\textrm{M}_{\odot} < \textrm{M}_{\star} < 1.5 \times10^{11}~\textrm{M}_{\odot}$.

\begin{figure*}
\begin{center}
\includegraphics[width=0.97\textwidth]{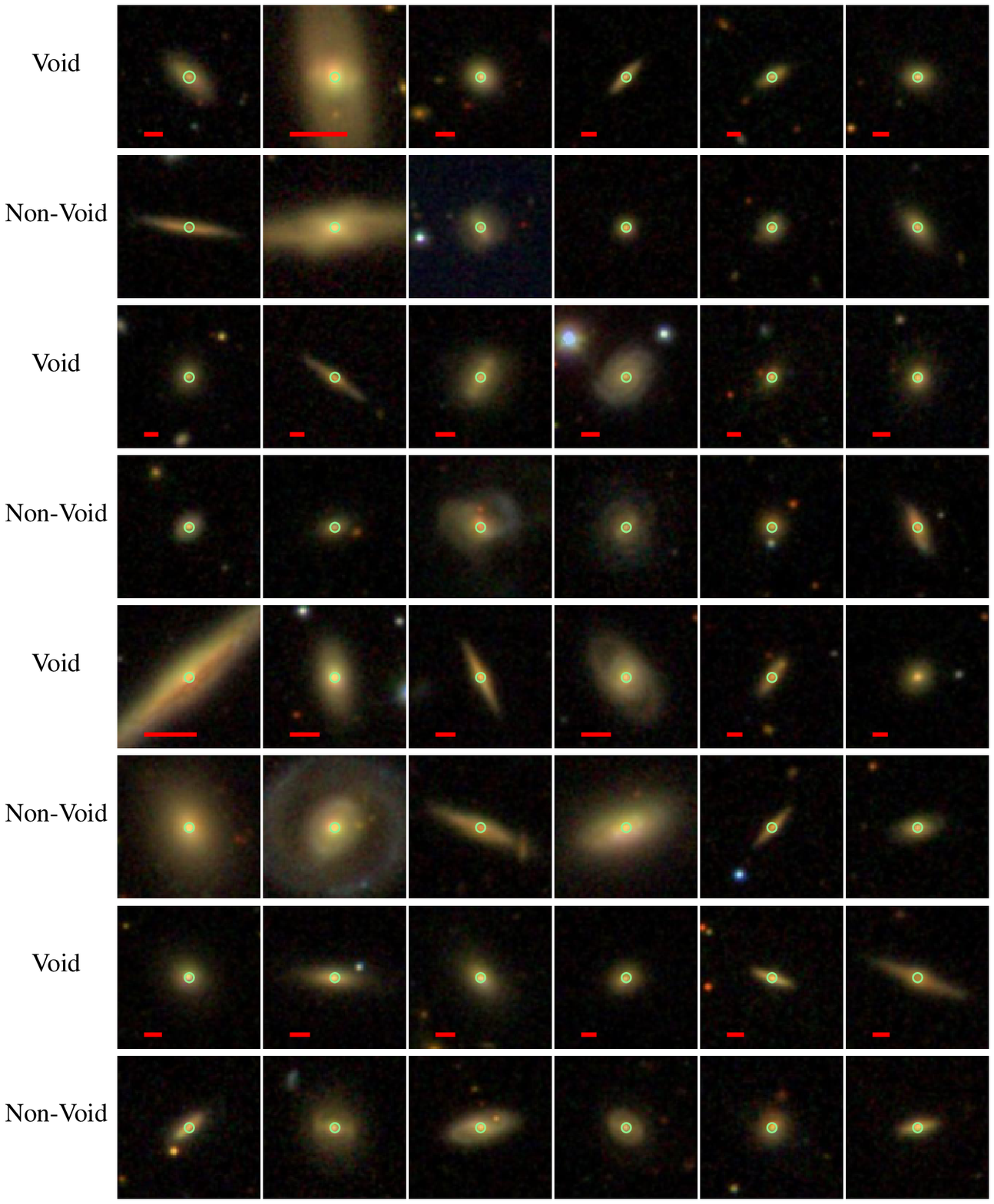} 
\end{center}
\caption{Void galaxies and a mass-matched random comparison galaxy selected to have similar properties to the void galaxy in terms of redshift and colour. The galaxies are sorted in order of ascending mass. The green circle in each plot is the SDSS single fibre spectroscopy coverage (3~arcsec diameter), and the red bar is 5~kpc in length. The galaxies are ordered from from $3.57\times10^{10}$~M$_{\odot}$ (top left) to $1.53\times10^{11}$~M$_{\odot}$ (bottom right). The galaxies are located between $z=0.042$ and $z=0.10$, and have colours $ 1.70 < (u-r) < 2.57$. It can be seen that a large fraction of massive void galaxies have disc-like morphologies, with spiral arms, bars, and star formation in their outer regions.  \label{massive}}
\end{figure*}

\begin{figure*}
\begin{center}
\includegraphics[width=0.97\textwidth]{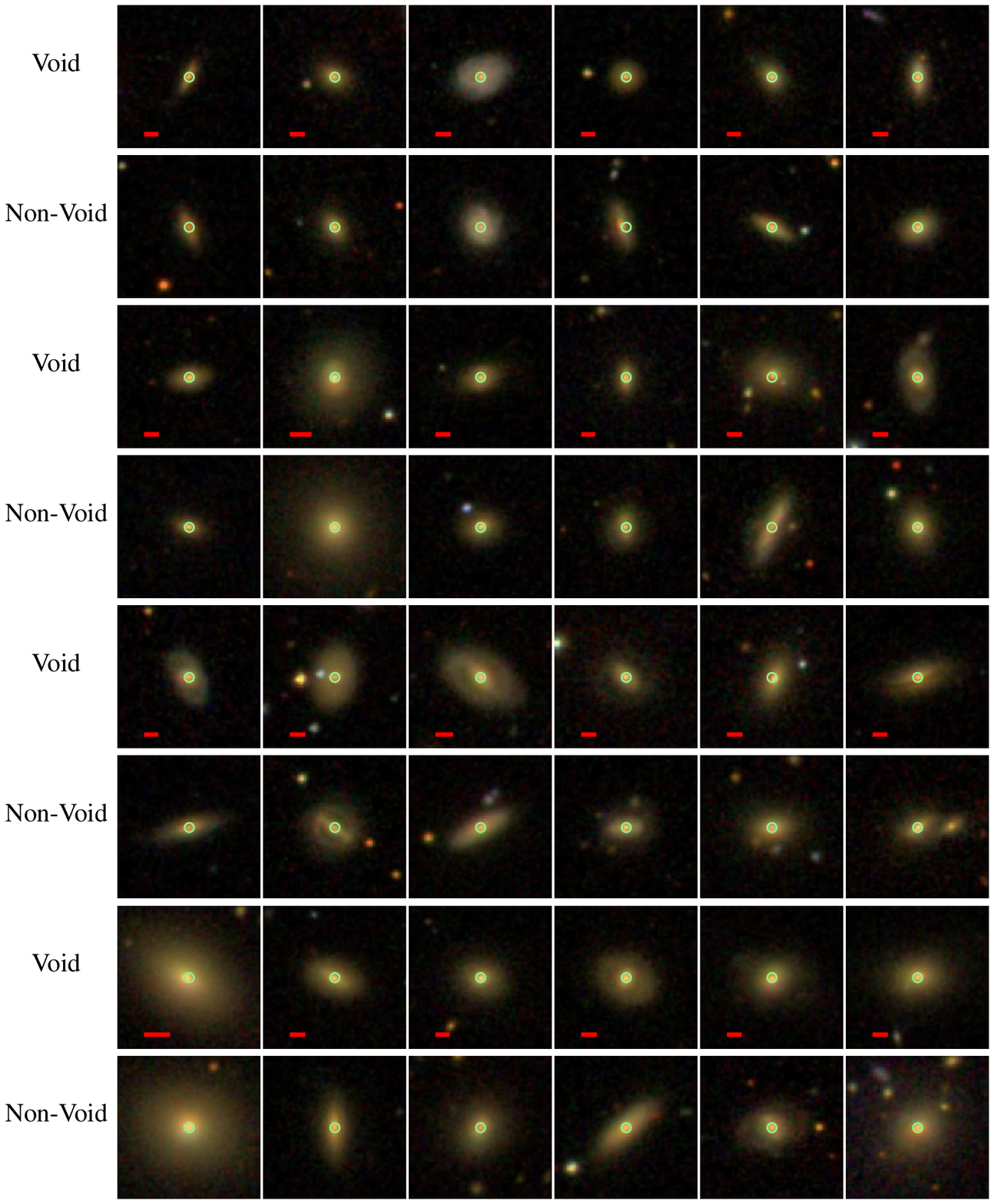}
\end{center}
\contcaption{}
\end{figure*}

The void galaxies exhibit a range of morphologies, however it is clear from Fig.~\ref{massive} that a large fraction have disc-like structures, with bar, ring, and spiral arm features common. 
A number of edge-on discs are seen, as well as ring, disc, and bar features in face-on galaxies that would be destroyed during a major merger. 
We hypothesise that based on their colours and morphologies, the highest mass void galaxies are red spirals and faded discs that have extinguished their gas supply, and have not undergone the interactions/mergers required to transform them into elliptical galaxies due to their isolation.

We furthermore compare the morphologies of the void galaxy population to a sample of 48 non-void comparison galaxies. 
For each void galaxy, a comparison galaxy was drawn from the non-void sample defined in Section~\ref{sec:fieldsamp}. 
The comparison galaxy was selected to have a similar mass (within 20~per~cent of the void galaxy's mass), rest-frame colour $(u-r) \pm0.15$, and redshift ($z\pm0.01$) to the void galaxy. 
The comparison galaxies are also shown in Fig.~\ref{massive}, underneath their corresponding non-void comparison galaxy. 
When the void and comparison samples are examined by eye, the morphologies of the two populations are similar, with the majority of the two samples being disc-dominated.  
To quantify these similarities, we examine the light distributions of the void and non-void comparison samples.

\subsection{Light distributions and bulge-disc decompositions}
\label{sec:decomp}

The bulge-to-disc ratio of a galaxy tells us much about its mass assembly history. 
A pure disc galaxy will have no bulge component, having formed its stellar population entirely via star formation. 
Likewise, a pure elliptical galaxy will have little/no disc component, having formed in a major merger, disrupting its discy component. 
Galaxies with structures intermediate between the two have a bulge+disc structure, and will have undergone a mixture of star formation, bar-drive evolution and merging, building up their bulge component, while maintaining or reassembling their disc. 
By examining the bulge/disc ratios of the void vs. non-void galaxies, we are able to see which process dominates the two galaxy populations. 

We expect the the impact of morphological transformation from disc-like to bulge-dominated via major mergers to increase in significance with increasing galaxy mass, such that bulge-to-total mass ratio and S\'ersic index $n$ will both increase as stellar mass increases \citep{2012MNRAS.421.1007K}. 
The most massive galaxies in the Universe are missing from voids (e.g. the brightest cluster galaxies and brightest group galaxies), so here we focus on the most massive void galaxies available, which only reach a stellar mass of $1.5\times10^{11}$~M$_{\odot}$, compared to masses $>5\times10^{11}$~M$_{\odot}$ expected for brightest cluster/group galaxies \citep[e.g.][]{2014MNRAS.440..762O}

To quantify the morphologies of the high mass GAMA void galaxy sample and examine their light distributions, we utilise the GAMA S\'ersic catalogue \citep{2012MNRAS.421.1007K}, and the SDSS bulge/disc decomposition catalogue of \citet{2011ApJS..196...11S}. 
The  bulge/disc catalogue provides three galaxy fitting models: a pure S\'ersic model, an $n_{b} = 4$ bulge + disc model, and a S\'ersic (free $n_{b}$) bulge + disc model for 1.12 million galaxies in the SDSS Legacy Survey. 
For the purpose of this paper, we utilise the  $n_{b} = 4$ bulge + disc mode to examine bulge-to-total stellar mass ratios.  

We analyse the light distributions of the void and non-void comparison galaxies using the GAMA S\'ersic photometry catalogue, with an update to this catalogue for GAMA~\textsc{ii} provided by \citet{2014arXiv1411.6355L}. 
We compare the S\'ersic indices of the void and comparison samples in the $r$-band for galaxies with $M_{\star} >5\times10^{9}$~M$_{\odot}$.
The mass-matched randomly drawn comparison sample of 304 galaxies was drawn from the non-void sample using the same criteria as the Section~\ref{sec:whancol}.

Clear trends are found for the S\'ersic indices of both the void and comparison sample as a function of mass. 
Below a stellar mass of $3\times10^{10}$~M$_{\odot}$, both samples favour low values of S\'ersic $n$, with a median S\'ersic $n= 1.27 \pm 0.12$ for the void population, compared to $n=1.39\pm0.10$ for the non-void population.  
For galaxies more massive than $3\times10^{10}$~M$_{\odot}$, we find $n=3.64 \pm 0.22$ for the void galaxy population, and $n=3.68\pm0.32$ for the non-void comparison sample. 
Within the errors, no difference is found for the light distributions of the void and non-void comparison samples.

We  examine the bulge/disc decompositions of our samples using the $r$ band $n_{b} = 4$ bulge + disc model of \citet{2011ApJS..196...11S}. 212 void galaxies with mass $>5\times10^{9}$~M$_{\odot}$ are present in the catalogue of \citet{2011ApJS..196...11S}, and a mass-matched randomly non-void comparison sample is drawn with the same sample size. Below a stellar mass $3\times10^{10}$~M$_{\odot}$, $3\times10^{10}$~M$_{\odot}$, both the void and comparison galaxies have a mean $g$-band bulge-to-total mass ratio $m_{bulge,g}/m_{total,g}=0.08$. 
Combined with low S\'ersic indices $\sim1$, in this mass regime, both the void and non-void populations are disc-dominated, having little/no bulge component, and have likely built up their mass via star formation and minor mergers.  

For galaxies with stellar masses $>3\times10^{10}$~M$_{\odot}$,  the $g$-band bulge-to-total mass ratios are near identical for the two population, with a median value $m_{bulge,g}/m_{total,g}=0.48\pm0.05$ for the void galaxies, vs. $0.54 \pm 0.05$ for the comparison sample. No obvious differences are found for the light distributions of the void and comparison samples over the entire mass range examined here.

\section{How do void galaxies acquire their mass? The major-merger histories of void galaxies from simulations}
\label{sec:modelmergers}

\begin{figure*}
\includegraphics[width=0.48\textwidth]{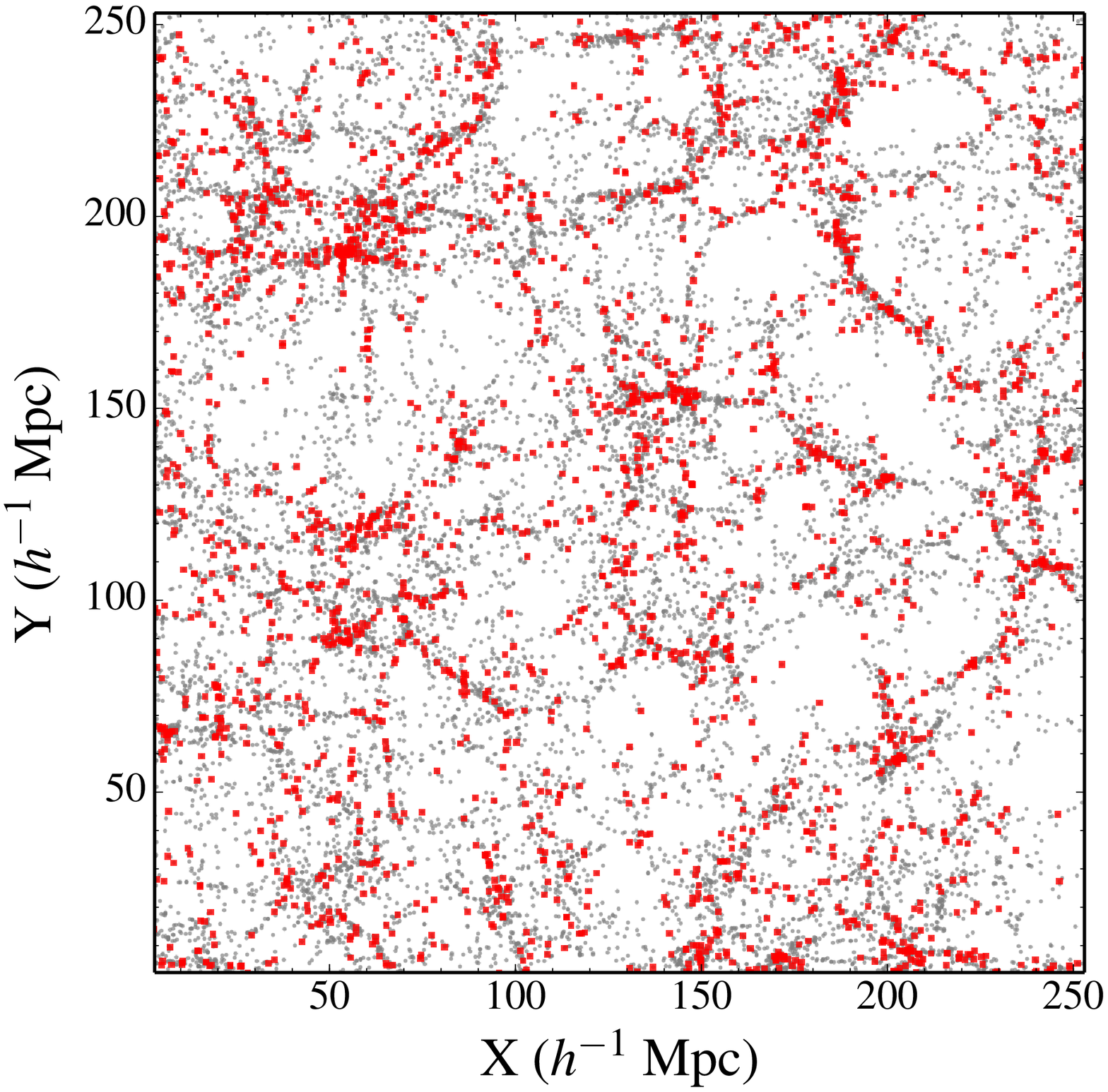}\includegraphics[width=0.48\textwidth]{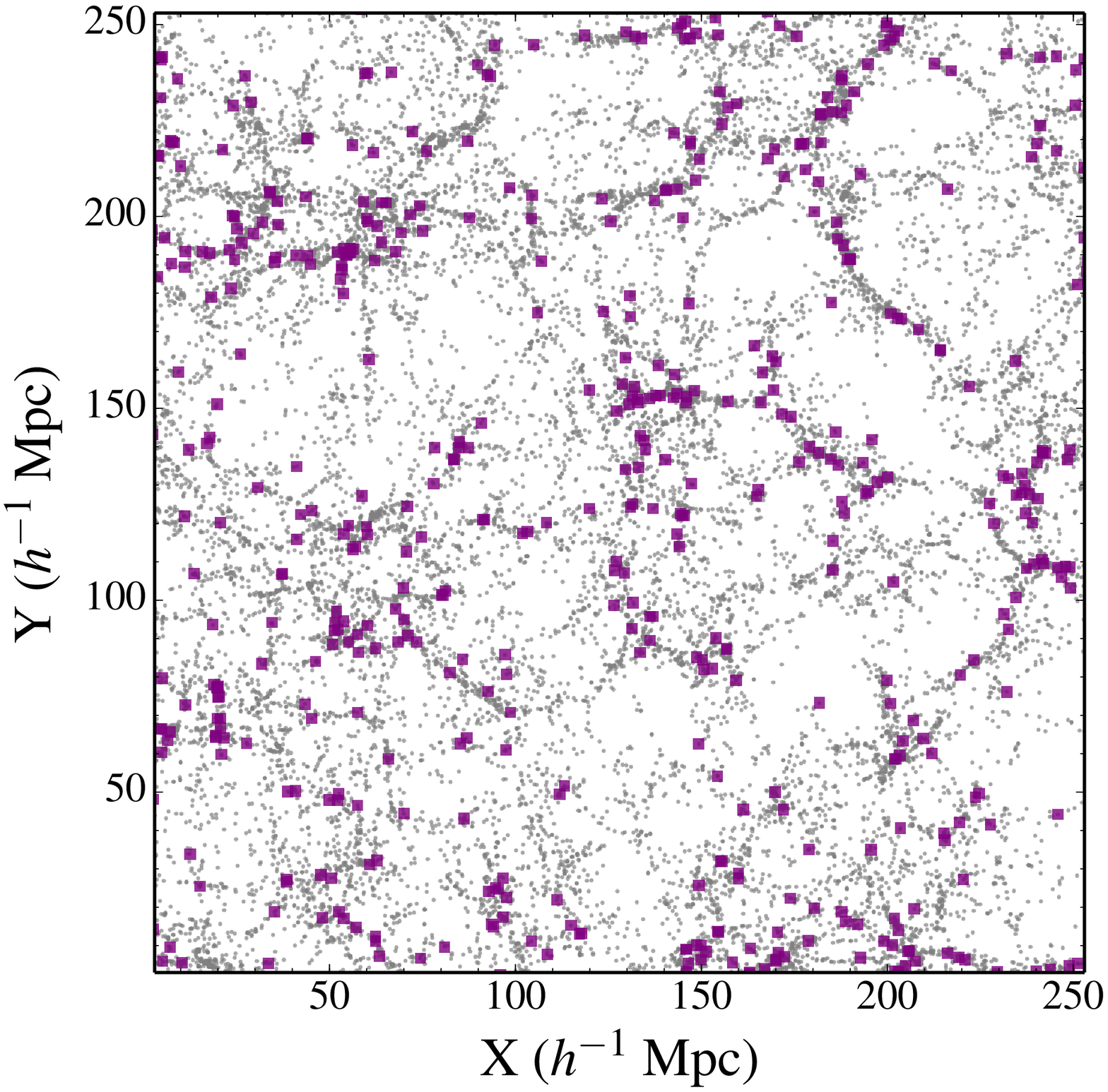}
\caption{Left panel: distribution of galaxies that have undergone a major merger at any point during their evolutionary histories (red points), taken from the Millennium Simulation. 
Right panel: the distribution of galaxies that have undergone major mergers in the past 5~Gyr only (purple points). In both plots, all galaxies in a $20~h^{-1}$~Mpc thick slice of the Millennium simulation at $z=0$ are shown as grey points. 
These grey points are common to both panels. 
The slice measures $250~h^{-1}~{\rm Mpc}\times250~h^{-1}~{\rm Mpc}$. The plotted galaxies have stellar mass $>5\times10^{9}$~M$_{\odot}$. Galaxies with major mergers in the past 5~Gyr typically avoid the void regions.}\label{mergers}
\end{figure*}

From Fig.~\ref{massive}, many of the highest mass galaxies and their non-void matched counterparts exhibit discy morphologies, with rings, bars, and spiral structures seen in many of the galaxies. 
 They furthermore have masses  $>3\times10^{10}$~M$_{\odot}$, the mass range where major mergers are hypothesised to have an increasingly important role in mass assembly. 
Unless the merger is sufficiently gas rich, major merger activity builds the spheroidal/bulge component of a galaxy, and destroys disc features, whereas mass assembly via star formation and minor mergers is less disruptive to the high mass progenitor's structure. 
Given they are typically disc dominated,  do massive void galaxies acquire their stellar mass via star formation or major mergers, and does the void merger fraction differ from that of the non-void comparison sample?  
As it is not possible to trace the mass assembly history of a galaxy, we must instead utilise results from N-body simulations, combined with semi-analytic models, to address this question. 
Rather than presenting observational results, the rest of this section uses these theoretical mass-assembly histories (via merger trees) to compare the major merger histories of void vs. non-void galaxies.
We utilise the Theoretical Astrophysical Observatory \citep{2014arXiv1403.5270B} to find the time since last major merger for galaxies in the Millennium Simulation \citep{2005Natur.435..629S}, separating galaxies by local galaxy density. 
The galaxies are modelled using the Semi-Analytic Galaxy Evolution (SAGE) galaxy model \citep{2006MNRAS.365...11C}. 
The code only outputs the time since the last major merger, though a galaxy may have undergone more than one major merger since its formation. 

To visualise how the major merger histories of void vs. non-void galaxies compare, we take a $250~h^{-1}~\textrm{Mpc}\times250~h^{-1}~\textrm{Mpc}\times20~h^{-1}$~Mpc thick slice from the original ($500~h^{-1}$~Mpc)$^{3}$, $z=0.0$ cube of the Millennium Simulation, and examine the distribution of galaxies that have undergone a major merger at any point during their evolutionary history. 
To these, we compare the distribution of those galaxies that have undergone a major merger in the past $5$~Gyr. 
The stellar mass range is selected to match that of the GAMA completeness limits of our observed galaxy samples. 

We quantify the local environment of the simulated galaxies using a 3rd nearest neighbour approach. 
A ($250~h^{-1}$)$^{3}$~Mpc volume of the $z=0$ Millennium Simulation containing 288354 galaxies with stellar masses $5\times10^{9}$~M$_{\odot}$ to $5\times10^{11}$~M$_{\odot}$ was utilised to find isolated void galaxies vs. non-isolated wall galaxies. 
The nearest neighbour search also included a 10~Mpc extension around the ($250~h^{-1}$)$^{3}$~Mpc volume, in case the third nearest neighbour lay outside the search volume.  
We use a third nearest neighbour distance $>6.3~h^{-1}$~Mpc to identify isolated galaxies. 
 
In Fig.~\ref{mergers}, we compare the distributions of all galaxies that have undergone major mergers (those having a mass ratio of 0.3 or higher between the merger progenitors) at some point in their history, vs. those that have undergone a major merger in the past 5~Gyr. 
Of the galaxies that have  undergone a major merger at some point in their evolutionary histories, 99.5~per~cent have 3rd nearest neighbour distances consistent with them lying in a cluster, group, or filament. 
For major merger in the last 5~Gyr (i.e. if most of their mass is assembled at recent times), 99.8~per~cent are found along regions of large scale structure. 
At all times, major mergers are preferably found outside of the void regions, but a small percentage of major mergers ($<1$~per~cent) take place in void regions.

For the mass range concerned, major mergers are relatively rare in both the void and wall galaxy populations. Approximately 21~per~cent of galaxies with stellar masses higher than $5\times10^{10}$~M$_{\odot}$ have undergone a major merger during their history, and this fraction drops to 10~per~cent for galaxies with masses between $5\times10^{9}$~M$_{\odot}$ and $5\times10^{10}$~M$_{\odot}$. 
Of the galaxies that have undergone major mergers, one void galaxy with a stellar mass $>5\times10^{10}$~M$_{\odot}$ has undergone a major merger in the past 5~Gyr, while $\sim11$~per cent of the wall galaxies with comparable mass have undergone major mergers in the past 5~Gyr. 

If we assume that major mergers at any past time are the means of morphological transformation from late-type discs to early-type ellipticals, we may expect  $\sim3$ of our void galaxies with masses $>5\times10^{10}$~M$_{\odot}$ to be early-types, and $\sim14$ non-void  mass-matched comparison galaxies to have early-type morphologies. 
If we assume major mergers within the last 5~Gyr are the means of morphological transformation (i.e. discs form around remnants for earlier mergers), then we may expect $\sim0$ of our void galaxies to be early-type and $\approx7$ non-void comparison galaxies to be early-types. 

Both major mergers and high mass galaxies are less common in isolation than in filaments, clusters, and groups. 
Star formation via gas accretion must be the dominant mass assembly process for the highest mass ($>5\times10^{10}$~M$_{\odot}$) isolated galaxies with third nearest neighbour distances $>6.3~h^{-1}$~Mpc, which explains the disc-dominated morphologies of the highest mass void galaxies shown in Fig.~\ref{massive}.
Major mergers are more common for galaxies in higher density regions of the Universe, and they are a significant mass assembly process for galaxies with third nearest neighbour distances $<1~h^{-1}$~Mpc and stellar masses $>5\times10^{10}$~M$_{\odot}$ i.e. massive galaxies in group scale environments or denser.

From the model and simulation, we conclude that a simple model of morphological transformation based on the presence/absence of major mergers in simulations predicts modest differences between the void and comparison galaxy samples matched in mass. 
However, we don't see evidence for this within our sample in Fig.~\ref{massive} nor in the quantitive measures of S\'ersic indices presented in Sec.~\ref{sec:decomp}.
Despite this lack of ellipticals in both our void and field observed samples, we cannot rule out the importance of major mergers in mass assembly. 
We expect the majority of elliptical galaxies to be found in clusters and massive galaxy groups. 
Only  7.5~per~cent of galaxies in the low-$z$ GAMA equatorial regions are found in groups with ten or more members, so our randomly-drawn non-void comparison sample could be lacking in elliptical galaxies.

\section{Discussion}
\label{sec:discuss}

Using the GAMA \textsc{ii} equatorial survey regions, we have identified a population of void galaxies to $z\approx0.1$ brighter than $M_{r} = -18.4$. 
At the mass range examined here, little to no difference is found between the void and non-void galaxy populations in terms of $(u-r)$ colour, \textit{WISE} mid-IR colours, star formation activity, line strength ratios, and morphology.
Our results are in close agreement with those of Alpaslan et al. (submitted), who find that isolated GAMA galaxies show a slight trend for higher IR emission, but have similar trends in ($u-r$) colour and morphology to galaxies in filaments. 

By comparing void with non-void comparison galaxies, we have found no evidence that quenching is dependent on large-scale environment. Quenched galaxies in voids overlap in morphology and mass in both samples (Fig.~\ref{irmass} and Fig.~\ref{massive}). 
This is consistent with quenching mechanisms that are a function of halo mass, such as virial shock heating, rather than environmental processes. 
We therefore find that galaxy mass is more important than environment in quenching star formation, again similar to the results of Alpaslan et al. (submitted). We discuss likely methods for mass quenching in the void sample, based on their morphologies, line strengths and mid-IR colours.

\subsection{Are optically red void galaxies really quenched?}
 
We classify the void galaxies according to their location on the WHAN diagram, to separate galaxies heated by star formation and nuclear activity, from those with old stellar populations. Using this diagnostic diagram, galaxies with a H$\alpha$ equivalent width $<3$~${\textrm {\AA}}$ have passive stellar populations, else are ionised by hot, evolved stars, rather than  AGN heating. \citet{2011MNRAS.413.1687C} state that both passive and retired galaxies have near-identical stellar populations, having retired from forming stars over $10^{8}$~yr ago. void galaxies have stellar populations inconsistent with either star formation or nuclear activity according to their classification on the WHAN diagram. 

However, when we examine the location of the passive/retired galaxies on the mid-IR colour-colour diagram, it is clear that a number of them must host optically obscured star formation, else are forming stars outside of the central regions sampled by single-fibre spectroscopy. 
A similar result is found using $(u-r) > 1.9$ as the colour cut for quenched vs. star forming stellar populations: the red void galaxies show a range of mid-IR $[4.6]-[12]$ colour inconsistent with them being a completely quenched galaxy population, though void galaxies with passive mid-IR colours are found. 
Current Integral Field Unit (IFU) spectroscopic surveys including SAMI \citep{2012MNRAS.421..872C} and MaNGA \citep{2015ApJ...798....7B} will allow us to examine the nature of these galaxies in more detail.  
These IFU surveys provide multiple spectra across the structure of a galaxy to $>1$~$R_{e}$, allowing us to characterise the ages of galaxies beyond their nuclear regions.

\subsection{Mass quenching in void galaxies}

Do void galaxies undergo mass quenching?  
To address this question, we use the $[4.6]-[12]$ mid-IR colours of the void galaxies as a proxy for current star formation activity.  
Similar to \citet{2011ApJ...735..112J} and \citet{2014ApJ...782...90C} we use $[4.6]-[12] = 1.5$ as the division between star forming and passively evolving stellar populations. 
In Fig.~\ref{irmass}, we bin the galaxies by mass, and examine how the distribution of galaxies on the $[3.4]-[4.6]$ vs. $[4.6]-[12]$ colour-colour diagram changes with mass. 
For both the void and non-void galaxy populations, the fraction of galaxies with passive $[4.6]-[12]$ colour increases with increasing galaxy mass, and passive void galaxies have stellar masses $>10^{10}$~M$_{\odot}$.

The GAMA galaxy group catalogue \citep{2011MNRAS.416.2640R} reveals these intermediate/high mass void galaxies with $M_{\star} > 10^{10}$~M$_{\odot}$ to be isolated, else they are the dominant/central galaxy in their halo (typically a pair or galaxy triplet). 
The isolated quenched void galaxies galaxies are unaffected by environmental processes such as strangulation and ram pressure stripping. 
Unless their gas supply has run out, internally driven evolutionary mechanisms must be responsible for the quenching of star formation activity in these void objects.

The quenched void galaxies in our sample have masses $\sim10^{10}$~M$_{\odot}$, and such galaxies typically occupy dark matter haloes with $M_{halo} > 10^{12}$~M$_{\odot}$ \citep[e.g.][]{2013ApJ...770...57B}. 
 This halo mass of $10^{12}$~M$_{\odot}$ is critical in models of feedback processes such as virial shock heating and AGN feedback. \citet{2008MNRAS.386.2285C} examine the halo mass function for both the void and general galaxy population in the 2dF Galaxy Redshift Survey. 
 They find that optically defined red sequence galaxies occupy the most massive haloes in all environments, and red void galaxies simply occupy the most massive dark matter haloes in voids. 
 They find that transition from optically blue-to-red dominated dark matter haloes occurs at $M_{vir} \sim 10^{12}$~M$_{\sun}$ for all environments, with halo mass the key driver in quenching star formation. 
 
Given their isolation, and therefore lack of environmental quenching mechanisms, void galaxies in halo masses $<10^{12}$~M$_{\odot}$ will be star forming. For these less massive haloes, gas is accreted with temperature $10^{4}$-$10^{5}$~K, and this gas collapses to form new stars.   
More massive than this critical dark matter halo mass of $10^{12}$~M$_{\odot}$, infalling gas is heated by shocks to near the virial temperature of $10^{6}$~K, and stars can no longer form \citep{2006MNRAS.368....2D}. 
The highest mass void galaxies we identify with stellar masses $>10^{10}$~M$_{\odot}$ reside in dark matter haloes with masses sufficient for this process to become efficient, and thus virial shock heating may prevent gas cooling and the formation of new stars.
Processes such as photoheating from the UV background may have suppressed star formation in  void dwarf galaxies \citep{2006MNRAS.371..401H}, but we are unable to identify such galaxies in this work due to incompleteness for low mass galaxies. 

Furthermore, feedback by AGNs is hypothesised to become efficient in dark matter haloes with masses $>10^{12}$~M$_{\odot}$.  
This AGN feedback preferentially affects the shock heated medium, and prevents it from ever cooling to form stars \citep{2006MNRAS.368....2D}.  
\citet{2008MNRAS.386.2285C} suggest radio-mode low-luminosity AGN heating as a possible source of the shut-down of star formation for passive void galaxies. 
The \textit{WISE} colours of radio-weak AGN are indistinguishable from those of passive/retired galaxies \citep{2014MNRAS.438.1149G}, and their spectra may not show signs of nuclear activity (their stellar component dominates the spectra) 
We are unable to identify such galaxies via the WHAN diagram or mid-IR colours (Fig.~\ref{whan} and Fig.~\ref{wisewhan}). 
We therefore consider weak radio-mode AGN heating as another possible quenching mechanism for the passive void galaxies. 
However, with our existing data, we are unable to confirm if these low-luminosity AGN exist in our sample, though several void galaxies do meet the mid-IR AGN classification criteria of \citet{2011ApJ...735..112J}. 

The majority of void and comparison galaxies in our sample exhibit discy morphologies in Fig.~\ref{massive}, so we also consider the possibility that minor mergers are responsible for the mass accretion, and eventual quenching, of the void galaxies. 
Minor mergers may play an important role in quenching star formation, while still preserving discy galaxy morphology \citep{2007MNRAS.380..339B}. 
The smooth accretion of material over time via minor mergers and gas accretion leave the disc of a galaxy intact, while building galaxy mass until the critical halo mass of $10^{12}$~M$_{\odot}$ for virial shock heating is reached.  
This is followed by bursty star formation and eventual quenching. 
This process can give rise to red and dead late-type galaxies in low mass, isolated haloes, with extended shut-down periods that last 6-7~Gyr. 

The star formation we detect in the mid-IR colour-colour diagram in Fig.~\ref{wisewhan} for optically passive void and non-void comparison galaxies could indeed be due to these isolated galaxies shutting down gradually over several Gyr. 
A combination of minor mergers and gas accretion is therefore a likely quenching mechanism for void galaxies, though a detailed examination of their star formation histories is required to confirm this.

\subsection{Mass-build up in void galaxies}

We show in Figure~\ref{massive} that the most massive void galaxies have disc-like morphologies, similar to their non-void counterparts. 
These disc-like structures suggest that gas accretion and minor mergers have been responsible for the mass assembly histories of these isolated galaxies. 
\citet{2004ApJ...617...50R} state that void galaxies are bluer than those in richer environments as those in clusters and groups have had their gas supply shut-off. 
If the void galaxies are able to sustain their gas supplies over a longer period of time than those in high-density regions, the void galaxies will continue to form stars after those in groups and clusters have been quenched \citep{2005ApJ...624..571R}.
\citet{2000AJ....119...32G} suggest that luminous galaxies at higher density have less gas and dust available for star formation than their void counterparts, due to their earlier formation timescale.
As a consequence of this difference in gas/dust content, luminous galaxies in voids that undergo interactions with nearby companions (e.g. minor mergers) are more able to have star formation triggered by such an encounter vs. those in groups and clusters, and will be more actively forming stars at the present time.

Evidence for this sustained gas accretion in void galaxies is found by \citet{2010ApJ...725.1550C}, \citet{2011AJ....141....4K} and \citet{2012AJ....144...16K} who show that many isolated/void galaxies have large reservoirs of H\textsc{i} gas, which favours secularly driven over environmentally driven evolution. 
Unlike their counterparts in regions of higher density, void galaxies are able to retain these large gas reservoirs that would otherwise be stripped due to environmental processes such as mergers.

Gas accretion as the primary driver of mass growth is furthermore supported by data from simulations, where major mergers are rare in isolated galaxies vs. wall galaxies found in filaments, groups and voids (Section~\ref{sec:modelmergers}). 
This comparison shows that 11~per~cent of the simulated wall galaxies have undergone a major merger in the past 5~Gyr, vs. just 2~per~cent of the simulated void galaxies.

While the transformation of discs into spheroids via major mergers is well accepted, major mergers may result in disc dominated galaxies provided the merging galaxies are sufficiently gas rich \citep[e.g.][]{2009MNRAS.397..802H}. 
Thus in a void or indeed any environment  where environmental processes are weak, galaxies will retain their gas, and more gas-rich than gas-poor major mergers might be expected. 
Major mergers may therefore play a role in mass-quenching void galaxies, though interactions between similar mass galaxies in voids are expected to be rare \citep[though see][]{2013AJ....145..120B}. 
The models of \citet{2013MNRAS.433.1479H} show that roughly 45\% of present day isolated galaxies have undergone at least one merger, and minor mergers account for 2/3 of these (with similar fractions found for galaxies not in isolation). 
Thus minor mergers are more important than major mergers for the mass assembly of isolated galaxies. 
We therefore consider major mergers as a relatively un-important method of mass build-up in void galaxies compared to minor mergers and gas accretion.

\section{Conclusions}
\label{sec:conclude}

\begin{itemize}
  \item We identify void galaxies with masses $>5\times10^{9}$~M$_{\odot}$ and $(u-r)$ colours consistent with those of a passively evolving stellar population. The WHAN line strength diagram reveals the majority of these red void galaxies to have ceased forming stars $>100$~Myr ago. 
  \item The mid-IR $[3.4]-[4.6]$ vs. $[4.6]-[12]$ colour-colour diagram reveals optically selected red-sequence void galaxies  to exhibit a range of $[4.6]-[12]$ colour, showing a wide spread in their recent star formation activity. 
  \item Regardless of environment, a trend is seen in mid-IR colour such that higher mass galaxies are bluer than lower mass galaxies. Mass quenching occurs in the void galaxy population, and massive galaxies will rarely form stars regardless of their global environment.  
  \item All void galaxies with passive mid-IR colours $[4.6]-[12]<1.5$ have stellar masses $>10^{10}$~M$_{\odot}$. A similar trend is seen for the non-void comparison galaxies. Regardless of global environment, central galaxies in sufficiently large haloes have passive stellar populations. 
  \item SDSS imaging reveals similar morphologies for the massive void and non-void galaxies at a similar mass, and the two populations have similar light distributions as indicated by their S\'ersic indices. The massive void galaxies are primarily discy in nature, as expected for galaxies in a low density environment.   
\end{itemize}

\section*{Acknowledgements}

SJP acknowledges the support of an Australian Research Council Super Science Postdoctoral Fellowship grant FS110200047, and postdoctoral funding from the University of Portsmouth. MB acknowledges financial support from the Australian Research Council (FT100100280) and the Monash Research Accelerator Program (MRA). We thank the reviewer for their comments which have improved this paper. GAMA is a joint European-Australasian project based around a spectroscopic campaign using the Anglo-Australian Telescope. The GAMA input catalogue is based on data taken from the Sloan Digital Sky Survey and the UKIRT Infrared Deep Sky Survey. Complementary imaging of the GAMA regions is being obtained by a number of independent survey programs including GALEX MIS, VST KiDS, VISTA VIKING, WISE, Herschel-ATLAS, GMRT and ASKAP providing UV to radio coverage. GAMA is funded by the STFC (UK), the ARC (Australia), the AAO, and the participating institutions. The GAMA website is http://www.gama-survey.org/.

Data used in this work was generated using the Theoretical
Astrophysical Observatory (TAO). TAO is part of the Australian All-Sky Virtual
Observatory (ASVO) and is freely accessible at https://tao.asvo.org.au.
The Millennium Simulation was carried out by the Virgo Supercomputing
Consortium at the Computing Centre of the Max Plank Society in Garching. It is
publicly available at http://www.mpa-garching.mpg.de/Millennium/. 



\bibliographystyle{mnras}
\bibliography{spenny} 





\bsp	
\label{lastpage}
\end{document}